\documentclass[12pt]{aastex}
\usepackage{emulateapj5}

\usepackage{graphicx}
\usepackage{lscape}

\newcommand{\etal}{\mbox{et~al.}}
\newcommand{\eg}{e.g.\@~}

\newcommand{\ie}{i.e.\@~}
\newcommand{\fig}{Fig.~}
\newcommand{\tbl}{Table~}
\newcommand{\eqn}{Eqn.~}
\newcommand{\inprep}{in prep.}

\shorttitle{SEDs of Local (U)LIRGs}
\shortauthors{U \etal}

\begin{document}

  \title{Spectral Energy Distributions of Local Luminous 
    and Ultraluminous Infrared Galaxies}
  
  \author{Vivian U\altaffilmark{1,2,3}, 
D. B. Sanders\altaffilmark{1}, 
J. M. Mazzarella\altaffilmark{4}, 
A. S. Evans\altaffilmark{5,6},
J. H. Howell\altaffilmark{4,7}, 
J. A. Surace\altaffilmark{7}, 
L. Armus\altaffilmark{7}, 
K. Iwasawa\altaffilmark{8}, 
D.-C. Kim\altaffilmark{5,6}, 
C. M. Casey\altaffilmark{1}, 
T. Vavilkin\altaffilmark{9}, 
M. Dufault\altaffilmark{10}, 
K. L. Larson\altaffilmark{1}, 
J. E. Barnes\altaffilmark{1},
B. H. P. Chan\altaffilmark{4}, 
D. T. Frayer\altaffilmark{11}
S. Haan\altaffilmark{12}, 
H. Inami\altaffilmark{13},
C. M. Ishida\altaffilmark{1}, 
J. S. Kartaltepe\altaffilmark{13}, 
J. L. Melbourne\altaffilmark{14}, 
A. O. Petric\altaffilmark{7} 
}

 \altaffiltext{1}{Institute for Astronomy, University of Hawaii, 2680
   Woodlawn Drive, Honolulu, HI 96822, USA; vivian@ifa.hawaii.edu}  
    %sanders,klarson,barnes,cmcasey@ifa.hawaii.edu,cmishida@mac.com}
 \altaffiltext{2}{NASA Jenkins Predoctoral Fellow}
 \altaffiltext{3}{SAO Predoctoral Fellow, Harvard-Smithsonian Center
   for Astrophysics, Cambridge, MA 02138, USA} 
  \altaffiltext{4}{Infrared Processing and Analysis Center, California
    Institute of Technology, 1200 E. California Blvd., Pasadena, CA 91125, USA} 
    %mazz@ipac.caltech.edu}
  \altaffiltext{5}{Department of Astronomy, University of Virginia,
    530 McCormick Road, Charlottesville, VA 22904, USA} 
    %aevans@virginia.edu}
  \altaffiltext{6}{National Radio Astronomy Observatory, 520 Edgemont
    Road, Charlottesville, VA 22903, USA} 
    %dkim@nrao.edu}
  \altaffiltext{7}{Spitzer Science Center, 1200 E. California Blvd.,
    MS 314-6, California Institute of Technology, Pasadena, CA 91125, USA}
    %jhhowell,jason,lee,bchan@ipac.caltech.edu,ap@astro.caltech.edu}
  \altaffiltext{8}{ICREA and Institut del Ci\`{e}ncies del Cosmos,
    Universitat de Barcelona (IEEC-UB), Mart\'{i} i Franqu\`{e}s, 1, 
    08028 Barcelona, Spain}  
    %kazushi.iwasawa@icc.ub.edu}
 \altaffiltext{9}{Department of Physics and Astronomy, Stony
   Brook University, Stony Brook, NY 11794, USA}   
    %tvavilk@vulcan.ess.sunysb.edu}
  \altaffiltext{10}{Department of Astronomy, Yale University, P.O. Box
    208101, New Haven, CT 06511, USA}  
    %dufault@astro.yale.edu}
  \altaffiltext{11}{National Radio Astronomy Observatory, P.O. Box 2,
    Green Bank, WV 24944, USA} 
    %dfrayer@nrao.edu}
  \altaffiltext{12}{CSIRO Astronomy and Space Science, Marsfield NSW,
    2122, Australia} 
    %sebastian.haan@csiro.au}
  \altaffiltext{13}{National Optical Astronomy Observatory, 950
    N. Cherry Ave., Tucson, AZ 85719, USA} 
    %inami,jeyhan@noao.edu}
  \altaffiltext{14}{Caltech Optical Observatories, Division of
    Physics, Mathematics \& Astronomy, MS 320-47, California Institute
    of Technology, Pasadena, CA 91125, USA} 
    %jmel@caltech.edu}

  \begin{abstract}
    Luminous (LIRGs: log~$(L_{\rm IR} / L_\odot) = 11.00 - 11.99$) and
    ultraluminous infrared galaxies (ULIRGs: log~$(L_{\rm IR} /
    L_\odot) = 12.00 - 12.99$) are the most extreme star forming
    galaxies in the universe. The local (U)LIRGs provide a unique
    opportunity to study their multi-wavelength properties in detail
    for comparison to their more numerous counterparts at high redshifts.
    We present common large aperture photometry at radio through X-ray
    wavelengths, and spectral energy distributions (SEDs) for a sample
    of 53 nearby ($z < 0.083$) LIRGs and 11 ULIRGs spanning
    log~$(L_{\rm IR} / L_\odot) = 11.14 - 12.57$ from the flux-limited
    ($f_{60 \mu m} > 5.24\,$Jy) Great Observatories All-sky LIRG
    Survey (GOALS).  The SEDs for all objects are similar in 
    that they show a broad, thermal stellar peak ($\sim 0.3-2\mu$m)
    and a dominant FIR ($\sim 40-200\mu$m) thermal dust peak, where
    $\nu L_\nu ({\rm 60\mu m}) / \nu L_\nu(V)$ increases from $\sim 2-30$
    with increasing $L_{\rm IR}$.  When normalized at IRAS-60$\mu$m,
    the largest range in the luminosity ratio, 
    $R (\lambda) \equiv {\rm log} [\frac{\nu L_\nu (\lambda)}{\nu
      L_\nu (60\mu{\rm m})}]$, observed over the full sample is seen
    in the Hard X-rays ($HX = $ 2-10 keV), where $\Delta R_{HX} =
    3.73$ ($\bar R_{HX} = -3.10$). %, and far-ultraviolet ($FUV = 1580
%    \AA$),  where $\Delta R_{FUV} = 3.18$ ($\bar R_{FUV} = -1.88$).
    A small range is found in the Radio (1.4 GHz), $\Delta R_{\rm 1.4
      GHz} = 1.75$, where the mean ratio is largest, ($\bar R_{\rm 1.4
      GHz} = -5.81$). 
    Total infrared luminosities, $L_{\rm IR}(8-1000 \mu{\rm
      m})$, dust temperatures, and dust masses were computed from
    fitting thermal dust emission modified blackbodies to the
    mid-infrared (MIR) through submillimeter SEDs.  The new results
    reflect an overall $\sim$0.02 dex lower luminosity than the original
    \emph{IRAS} values. Total stellar masses were computed by 
    fitting stellar population synthesis models to the observed
    near-infrared (NIR) through ultraviolet (UV)  SEDs.  Mean stellar
    masses are found to be $\log(M_\star/M_\odot) = 10.79 \pm 0.40$.
    Star formation rates have been determined from the infrared
    (SFR$_{\rm IR} \sim 45 M_\odot$ yr$^{-1}$) and from the
    monochromatic UV luminosities (SFR$_{\rm UV} \sim 1.3 M_\odot$ yr$^{-1}$),
    respectively.  Multiwavelength AGN indicators have be used to
    select putative AGN: about 60\% of the ULIRGs would have been
    classified as an AGN by at least one of the selection criteria.  

  \end{abstract}
  

  \section{Introduction}
  \label{Introduction}

  Luminous infrared galaxies (LIRGs: $L_{\rm IR}[8-1000 \mu
  m]~\geq~10^{11}L_{\odot}$) are an important class of extragalactic
  objects.  Although relatively rare in the local universe, they still
  outnumber optically selected starburst and Seyfert galaxies at
  comparable bolometric luminosity \cite[][]{Soifer87}, and at the
  highest  luminosities, ultraluminous infrared galaxies, (ULIRGs:
  $L_{\rm IR}[8-1000 \mu m]~\geq~10^{12}L_{\odot}$),  exceed the space
  density of optically selected quasars by a factor of $\sim3$
  \cite[][]{Sanders88}.  Extensive follow-up observations at radio
  through X-ray wavelengths of complete samples of objects first
  discovered in the $IRAS$ All-Sky survey, show that (U)LIRGs appear to
  be powered by a mixture of starburst and AGN activity, triggered by
  strong interactions and mergers of gas-rich spirals \cite[see][for a
  more complete review]{Sanders96}.    
  
  \setcounter{footnote}{0}

  Despite the nearly two decades since the publication of the first
  $IRAS$ catalogs of (U)LIRGs, there is surprisingly little published
  photometry that can be used to construct accurate SEDs for even the
  nearest and best studied sources.  The majority of the nearby
  (U)LIRGs are often ``messy" systems that do not lend themselves to
  simple single aperture measurements.  Published data not only suffer
  from different apertures used at both the same and different
  wavelengths, but also from inconsistent definitions of the true
  extent and shape of the interacting galaxies, which are often
  characterized by highly irregular tidal debris fields. The Great
  Observatories All-sky LIRG Survey \cite[GOALS: ][]{Armus09} has made
  one of its top priorities, the compilation of a consistent set of
  photometric images of all LIRGs in the $IRAS$ Revised Bright Galaxy
  Sample \cite[RBGS: ][]{Sanders03}, by reanalyzing existing archival
  data and obtaining new images at radio through X-ray
  wavelengths\footnote{A complete description of the multi-wavelength
    data can be found on the GOALS website at http://www.goals.ipac.caltech.edu.}.
  
  This paper presents photometric radio through X-ray SEDs for 64 of
  the nearest and best studied (U)LIRGs, using common aperture
  ``masks" to compute accurate total fluxes (including the extended
  tidal debris fields) for each source.  These SEDs are then used to
  compute basic properties for each source --- the total infrared
  (IR)  luminosity, dust temperature, dust mass, and total stellar
  mass. We also compare SEDs of individual sources in 
  order to understand the expected variation in spectral shapes and
  colors as well as AGN diagnostics for the complete sample of
  (U)LIRGs.  Understanding the range of spectral and physical
  properties for this class of objects is critical before any direct
  comparison to their high-$z$ counterparts may be made.  
  
  The paper is organized as follows: descriptions of the (U)LIRG
  sample and of the multi-wavelength data sets are provided in \S
  \ref{Sample} and \S \ref{Data}, respectively.  The complete SEDs and
  spectral properties are presented in \S \ref{Results}. Derived
  and adopted properties, such as the infrared luminosity,  $L_{\rm
    IR}(8-1000\mu$m), dust temperature, dust mass, and stellar mass,
  $M_\star$, are given in  \S \ref{Discussion}. Our conclusions 
  are summarized in \S \ref{Summary}.  Throughout this paper, we adopt
  a flat model of the universe with a Hubble constant  $H_0 =
  70$\,km\,s$^{-1}$\,Mpc$^{-1}$, and  $\Omega_{\rm m}$ = 0.28,
  $\Omega_\Lambda$ = 0.72 \cite[][]{Komatsu09}. 

  \section{Sample}
  \label{Sample}

  Our sample of 64 (U)LIRGs represents the nearest and brightest 
  infrared-luminous extragalactic sources observable from  the
  northern hemisphere.  It is a complete subset of 
  all objects in the \emph{IRAS} Bright Galaxy Sample~\cite[][]{Soifer87}, 
  with $L_{\rm IR} > 10^{11.14} L_\odot$, originally chosen to satisfy the 
  constraints $|b| > 30^\circ$ and $\delta > -30^\circ$ in order to minimize 
  Galactic extinction and to be observable from Mauna Kea,
  respectively\footnote{NGC 1068 would have been a part of this sample.
  However, its proximity and thus, large angular extent, pose a
  challenge for the UH-2.2m telescope and its limited field-of-view to
  capture the entire galaxy and its full debris field. Therefore, it has
  been left out of the current sample.}.  
  Our subsample represents 30\% of all LIRGs and 50\% of all ULIRGs in GOALS.
  The median infrared luminosity of our subsample is $\log(L_{\rm
    IR}/L_\odot) = 11.60$,  with a luminosity range $\log(L_{\rm
    IR}/L_\odot) = 11.14 - 12.57$.   The redshift range is  $z =
  0.012 - 0.083$, corresponding to a luminosity distance, $D_L = 50.7
  - 387$ Mpc, with median $z = 0.028$ ($D_L = 119.0$ Mpc).  Basic
  properties of our subsample of GOALS objects are listed in \tbl
  \ref{tbl:basicinfo}.

  \section{Data}
  \label{Data}

   The multi-wavelength images for our targets have been assembled from 
   a variety of sources, including our own extensive archive of ground- and 
   space-based observations of (U)LIRGs, as well as data obtained from 
   various data archives.  The sources for the images used in our
   analysis is detailed in Tables \ref{tbl:newobs} and
   \ref{tbl:uhobs}.  All images were reprocessed and analyzed using a
   consistent set of criteria.   The procedures used to prepare the
   photometry masks are described in \S \ref{Masks}.   
 
   \subsection{Space-based Observations}
   \label{Space}
   
   Observations of the majority of our targets at X-Ray, ultraviolet,
   and infrared wavelengths were obtained as part of major GOALS observing 
   campaigns. The \emph{Chandra}-GOALS  program \cite[][]{Iwasawa11} provides 
   X-ray photometry  in the Soft X-ray (SX: 0.5-2 KeV) and Hard X-ray
   (HX: 2-10 KeV) bands. The \emph{GALEX}-GOALS
   program~\cite[][]{Howell10} provides observations of  the majority
   of our targets in the far-UV (FUV: $\lambda_{\rm eff} = 0.1528
   \mu$m) and near-UV (NUV: $\lambda_{\rm eff} = 0.2271 \mu$m)
   bands. The \emph{Spitzer}-GOALS programs (Mazzarella \etal~\inprep;
   Surace \etal~\inprep) obtained MIR (3.6, 4.6, 5.4, 8.0$\mu$m) and
   FIR (24, 70 and 160$\mu$m) images with IRAC and MIPS, respectively.
   The \emph{Herschel}-GOALS programs (PIs: Sanders, Armus) involving
   PACS and SPIRE observations are currently ongoing and will be
   completed in late 2012.  
   \emph{IRAS} data at 12, 25, 60, 100$\mu$m as published in the
   RBGS~\cite[][]{Sanders03} have also been incorporated into our
   analysis for completeness. 
  
   \subsection{Ground-based Observations}
   \label{Ground}
   
   Most of the observations in the optical bands were taken with the
   Tektronix 2048$\times$2048 camera (hereafter, Tek2048) at the
   University of Hawaii (UH) 2.2m Telescope on Mauna Kea.
   \emph{BVI}-band images for 53 LIRGs + 1 ULIRG were originally
   obtained as part of a PhD Thesis  that  studied a complete
   subsample of LIRGs from the \emph{IRAS} BGS~\cite[see][for details
   on the observations and reduction]{Ishida04}.  The primary source
   of the $R$-band images is the RBGS $R$-$K'$ Atlas (Mazzarella
   \etal~\inprep), while \cite{Surace98,Surace00a}, and
   \cite{Surace00b} provide $UBI$-band data for many of the ULIRGs,
   all imaged with either the Tek2048 or the Orbit 2048$\times$2048
   cameras on the UH 2.2m Telescope.  

  The remaining optical, near-infrared (NIR), submillimeter and radio data have been
   compiled from the literature and the NASA/IPAC Extragalactic Database
   (NED).  Most notably,  \emph{JHK$_S$} images were extracted from
   the 2-Micron-All-Sky-Survey \cite[2MASS;][]{Skrutskie06} Extended
   Source Image Server, and in the case of large systems spanning
   multiple frames, the Large Galaxy Atlas~\cite[][]{Jarrett03} via
   the InfraRed Science Archive (IRSA).  Submillimeter data at
   850$\mu$m and 450$\mu$m, obtained using the Submillimeter
   Common-User Bolometer Array (SCUBA) at the James Clerk Maxwell
   Telescope were taken from~\cite{Dunne00} and~\cite{Dunne01}, while 1.49
   GHz radio data obtained at the Very Large Array (VLA) were taken
   from~\cite{Condon90,Condon96}.   All of our archival data sources
   are summarized in Table \ref{tbl:newobs}. 

   The remaining gaps in our optical photometry table were filled by 
   observed with the Tek2048 camera on the UH 2.2m Telescope during 
   two observing runs, totaling 8 partially photometric nights
   in February and August of 2008.  Each exposure was typically 2-6
   minutes, with seeing ranging between $0\farcs5$ -- $0\farcs8$, (see
   \tbl \ref{tbl:uhobs} for a summary). A dither pattern of $80\farcs0
   \times 80\farcs0$ was used during the observations.  The data were
   reduced and calibrated using various 
   packages provided by the IDL astron library. The reduction involved
   several standard steps: all images were bias-subtracted using
   a median bias computed from bias frames taken at the beginning of 
   each night. There is a location-dependent shutter correction time
   of 0.18 seconds (Helene Courtois, private communication) for the 
   Tek2048; this is a significant contribution for the images with
   short exposure times (calibration stars in particular), so all of the
   data were corrected by this factor.  For each of the
   \emph{UVRI} filters during each night of the run, a master flat,
   created from median-combining normalized flats, was used to
   flat-field all the corresponding science images.  After
   adding astrometry information (using IDL routine STARAST) to
   the image headers, science images observed close in time within the
   same night were coadded to increase the signal-to-noise ratio for
   each target frame. The data were subsequently calibrated with
   standard stars in the corresponding optical
   bands~\cite[][]{Landolt83}, taking into account airmass
   corrections.  To ensure that consistent apertures were used to find
   the total flux from each galaxy system, photometry was subsequently
   measured using previously constructed ``masks'', as described in \S
   \ref{Masks}.

   \subsection{Photometry Masks}
   \label{Masks}

   When assembling SEDs for our sources we constructed a single
   photometric mask for each source that was designed to incorporate the
   total flux from a galaxy system whether it consists of a single
   galaxy with disturbed morphology or two separate interacting
   galaxies. The masks have been defined based on isophotes in the
   median- and boxcar-smoothed $I$-band images at the surface
   brightness limit of 24.5 mag/arcsec$^2$~(Vavilkin \etal~\inprep).
   They are intended to encapsulate the global flux from tidal debris
   as well as individual components within these merger systems. 
   The more luminous 43 objects in our sample 
   ($L_{\rm IR}\,>\,11.4$) have masks defined using \emph{HST}-ACS
   images (Evans \etal~\inprep); the other 21 masks were generated
   using the same technique from ground-based $I$-band images taken
   with the UH 2.2m Telescope. We have tested for systematic bias
   between the masks made from \emph{HST} and ground-based images and
   found that the difference in measured photometry is less than
   0.2\%, within photometric uncertainties.  The photometry masks for all 
   64 (U)LIRGs are shown in \fig \ref{fig:masks}. 

   Masked photometry has been extracted from images taken at effective
   wavelength $0.15 \mu$m$\,<\,\lambda_{\rm eff}\,<\,8 \mu$m and at MIPS
   24$\mu$m band.  The short wavelength limit has been imposed due to
   the fact that the X-ray photon counts are fairly localized, 
   whereas on the longward side, the images are lacking in resolution
   ($\gtrsim 6\farcs0$ PSF). At either end of the wavelength range,
   therefore, masks would not improve the precision of the total flux
   measurement over that of circular or elliptical apertures.
   Differential emission will be taken into account when deep
   optical/NIR images probing the extended debris field are acquired in
   the near future (\emph{Spitzer} snapshot program, P.I. Sanders). Within
   the wavelength regime where masks have been applied, the masks are
   very large relative to the beam size and hence aperture corrections
   are not needed.

  \section{Spectral Energy Distribution of the GOALS Sample}
  \label{Results}

\begin{figure}
    \centering
    \includegraphics*[width=0.99\textwidth,bb=0 320 612 792]{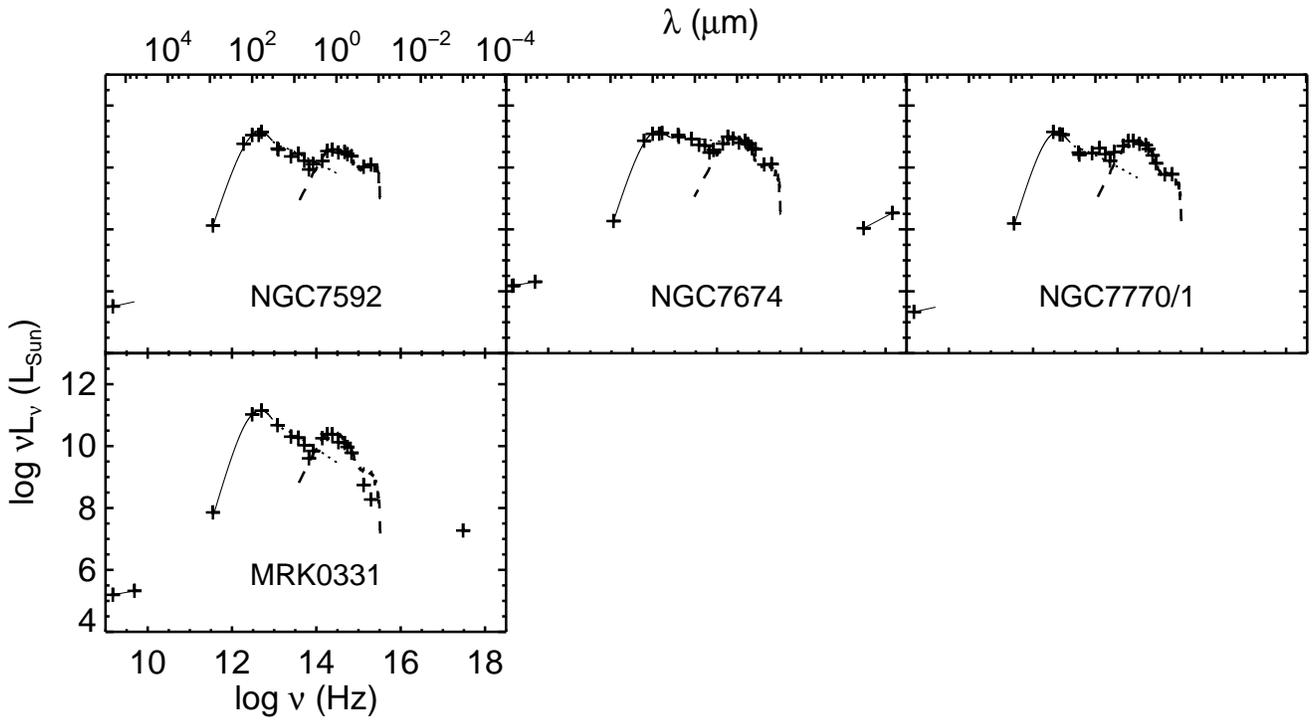}
    \caption{SEDs of the 64 objects in our sample in RA order. The
      units are $\log\,\nu$ (Hz) on the x-axis and $\log\,\nu L_{\nu}$
      ($L_\odot$) on the y-axis.  The crosses represent our photometry
      data points, while the dashed line illustrates the UV-NIR fit to
      \cite{Bruzual03} stellar population synthesis models. The mBB
      greybody fit composes of the mid-infrared powerlaw portion
      (dotted) and the far-infrared blackbody portion (solid) in the 
      SED.}
    \label{fig:allseds}
  \end{figure}

  In this section we present SEDs (radio through X-ray) for our complete 
  sample of 64 (U)LIRGs.
  The common aperture mask photometry data are provided in \tbl
  \ref{tbl:photometry1} (X-ray to NIR) and \tbl 
  \ref{tbl:photometry2} (MIR to radio).    The complete radio through X-ray
  SEDs ($\log \nu L_{\nu}$ vs $\log \nu$) for each source are shown in the 
  panels of  \fig \ref{fig:allseds}. The photometric data points are
  overlaid with the best-fit model SEDs, which will be discussed in
  more detail in \S \ref{Lephare}.  
  
   (U)LIRGs are known to exhibit several common features in their SEDs,
   \eg a drop-off in the UV flux, an optical-NIR stellar thermal
   ``bump", and a thermal dust ``bump" at FIR
   wavelengths~\cite[][]{Sanders96}.  However, photometry compiled
   for our complete local sample displays varied spectral
   shapes across the electromagnetic spectrum.  The main goal of our
   current  study is to characterize the full SED and spectral
   properties of (U)LIRGs as follows.
      
   \subsection{Spectral Shapes}
   \label{Templates}

   Before we employ the SEDs of these (U)LIRGs as local templates, we
   must first understand the variations in the spectral shapes across
   the range of various attributes.  In particular, \fig
   \ref{fig:seds_stack_nulnu} shows the radio through X-ray  SEDs of
   all 64 (U)LIRGs.  
   Normalized at $J$-band, this plot shows the best-fit modeled
   fluxes at the observed filters and brings out the variations in the optical
   (UV-NIR) regime, the shortward end of which at FUV is characterized
   by the young stellar populations in the galaxies.  No attempt has
   been made to connect the observed X-ray emission to the UV flux
   given our lack of data in the nearly two decade wavelength range
   between the FUV and soft X-ray (SX) measurements.  
   
\begin{figure}
    \centering
    \includegraphics[width=0.8\textwidth]{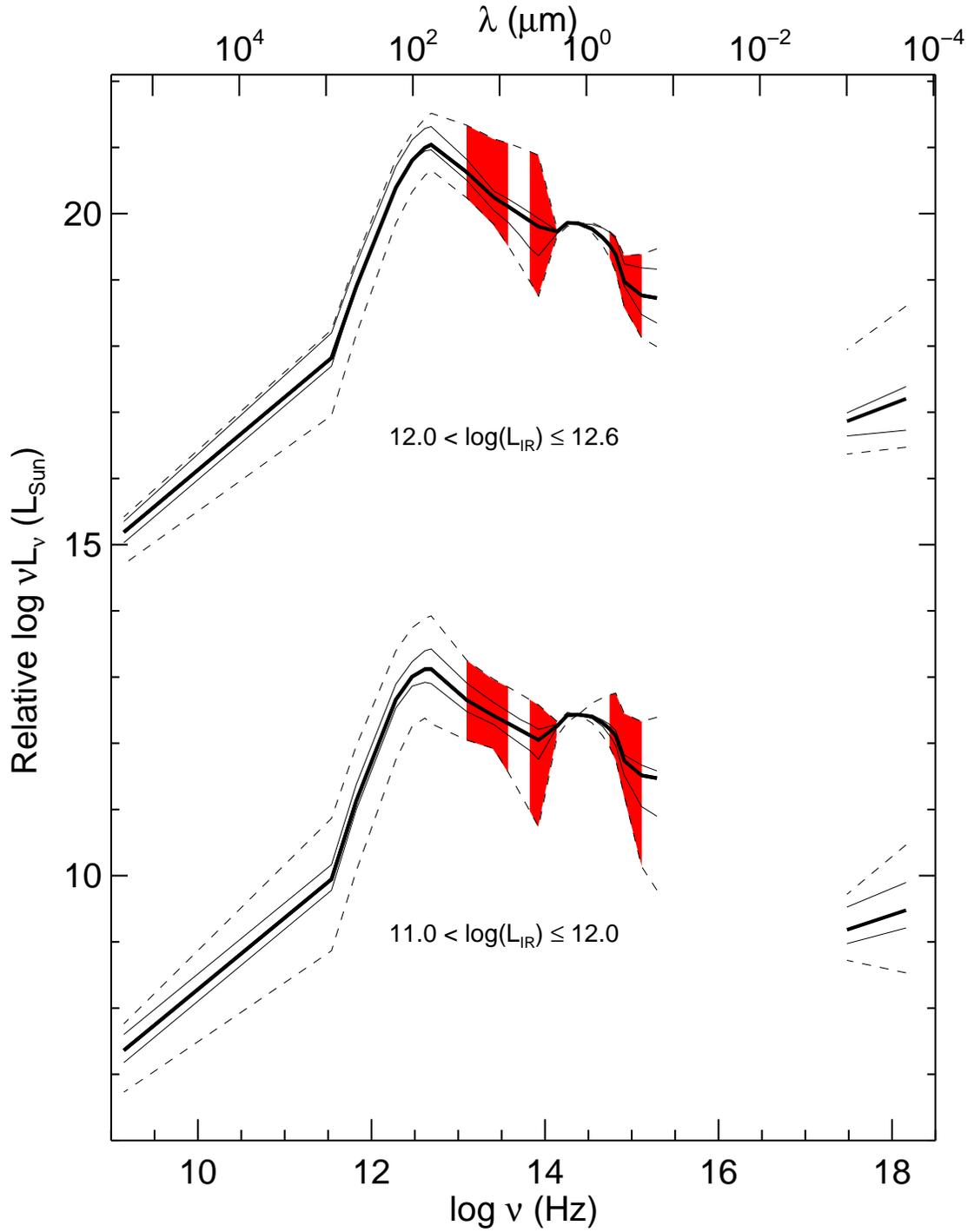}
    \caption{Mean radio through X-ray  SEDs normalized at $J$-band in
      two luminosity bins corresponding to ULIRGs (top) and LIRGs
      (bottom).  The 50\% and 100\% range for the SEDs are shown by
      the thin solid and dashed lines, respectively.  The shaded
      regions mark the wavelength ranges used to calculate the three
      spectral indices, $\alpha_2^1$, as described in the text. }  
    \label{fig:seds_stack_range}
  \end{figure}

   To better demonstrate both the range and mean of the
   object SEDs, we show in \fig \ref{fig:seds_stack_range} the mean
   SED along with the 25\%, 75\%, and max/min values for the 11 ULIRGs
   and 53 LIRGs separately. Qualitatively speaking, the overall SED shape
   is similar for all of the LIRGs and ULIRGs, with two significant
   maxima, one of which is near 1$\mu$m and the other near 80$\mu$m,
   with a large dip between them, and all SEDs falling towards the FUV
   and at submillimeter wavelengths. The logarithmic difference between
   the stellar and dust peaks is 1.2 dex for the ULIRGs but only 0.7
   dex for the LIRGs, providing a quantitative measure for the change
   in stellar-to-dust ratio with infrared luminosity. The next two
   sections offer different ways to quantify the spectral shapes.  
   The fit data values for the mean SEDs for both LIRGs and ULIRGs are
   given in \tbl \ref{tbl:templates}. 

 \subsection{Spectral Indices}
   \label{Indices}
\begin{figure}
    \centering
    \includegraphics[width=0.8\textwidth,angle=90]{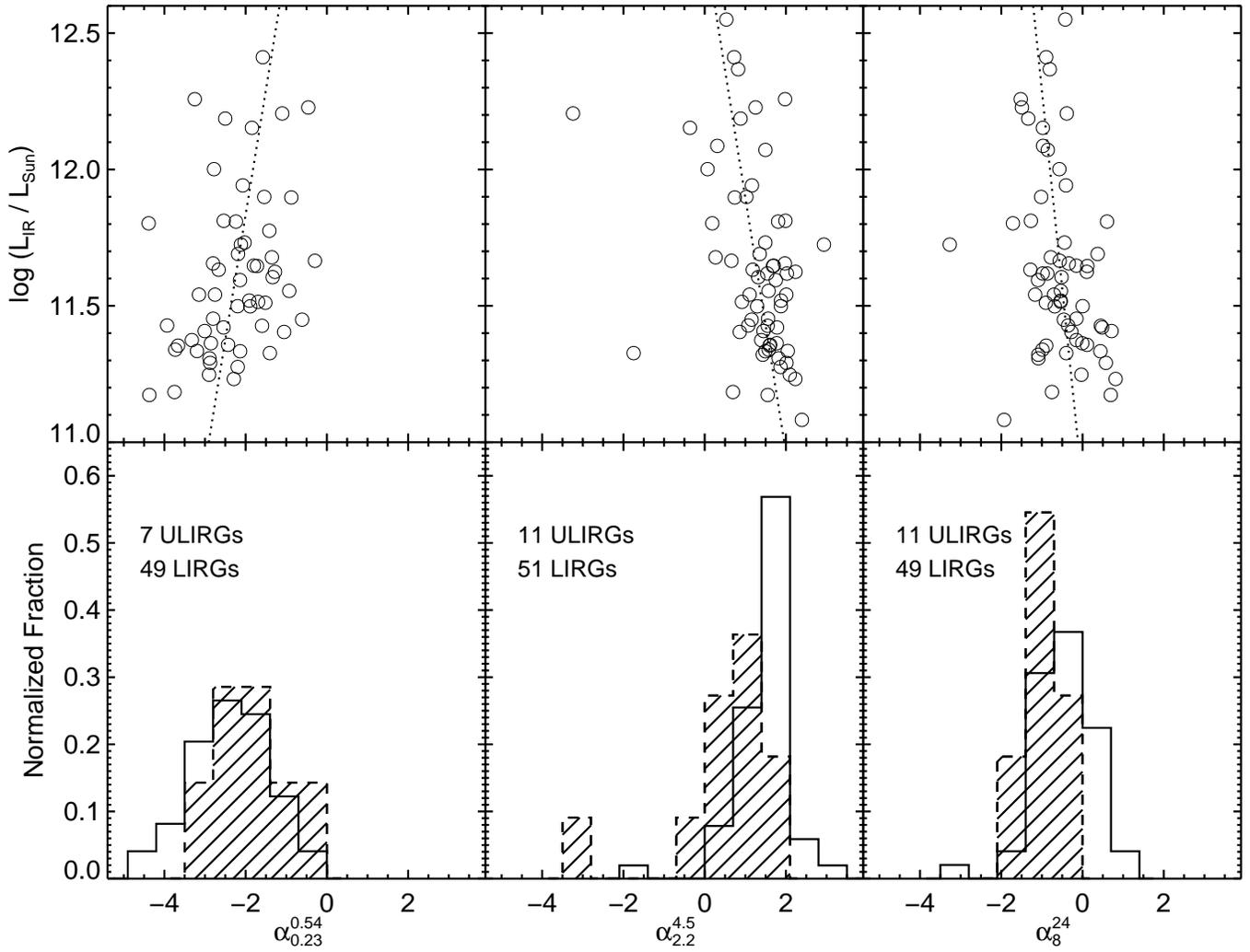}
 \caption{Spectral indices ($\alpha_{\rm 0.23}^{0.54}$, $\alpha_{\rm 2.2}^{4.5}$,
      and $\alpha_{\rm 8}^{24}$) plotted against infrared luminosity (top), and 
      their corresponding histograms (bottom). The best-fit (dotted)
      line within each of the top three panels shows the trends of the
      indices with increasing luminosity. The numbers of (U)LIRGs
      that are used to compute these indices (due to availability in
      the corresponding photometry data) are listed in the upper left-hand
      corner of the histograms. The hashed histograms represent ULIRGs
      while the open histograms represent LIRGs.} 
    \label{fig:alpha}
  \end{figure}

   A useful and more quantitative way of discussing the spectral
   shapes of SEDs makes use of spectral indices defined as follows:  
   \begin{equation}
     \alpha_2^1 = \frac{\log \nu_1 f_{\nu_1} - \log \nu_2
       f_{\nu_2}}{\log \nu_1 - \log \nu_2} \quad,
   \end{equation}
  where the indices (1,2) correspond to observed wavelengths in microns.  
  To mirror a high-$z$ SED study of a 70-$\mu$m seltected sample of
  (U)LIRGs~\cite[][]{Kartaltepe10},
   here we have chosen three wavelength ranges where the SEDs appear 
   to show the largest variations - in the UV-optical ($0.23 - 0.54 \mu$m), 
   NIR ($2.2 - 4.5 \mu$m) and the MIR ($8 - 24 \mu$m).  These three
   wavelength ranges correspond to the shaded regions shown in  \fig
   \ref{fig:seds_stack_range}. The top panels in \fig \ref{fig:alpha}
   show $\alpha_2^1$ vs. $L_{\rm IR}$ for three different spectral
   indices, corresponding to the three wavelength ranges described
   above.  The mean values of $\alpha_2^1$ for LIRGs and ULIRGs along
   with a regression analysis for the full subsample of 64 objects is
   given in \tbl \ref{tbl:alphas}.   
   
\begin{figure}
    \centering
    \includegraphics[width=0.8\textwidth,angle=90]{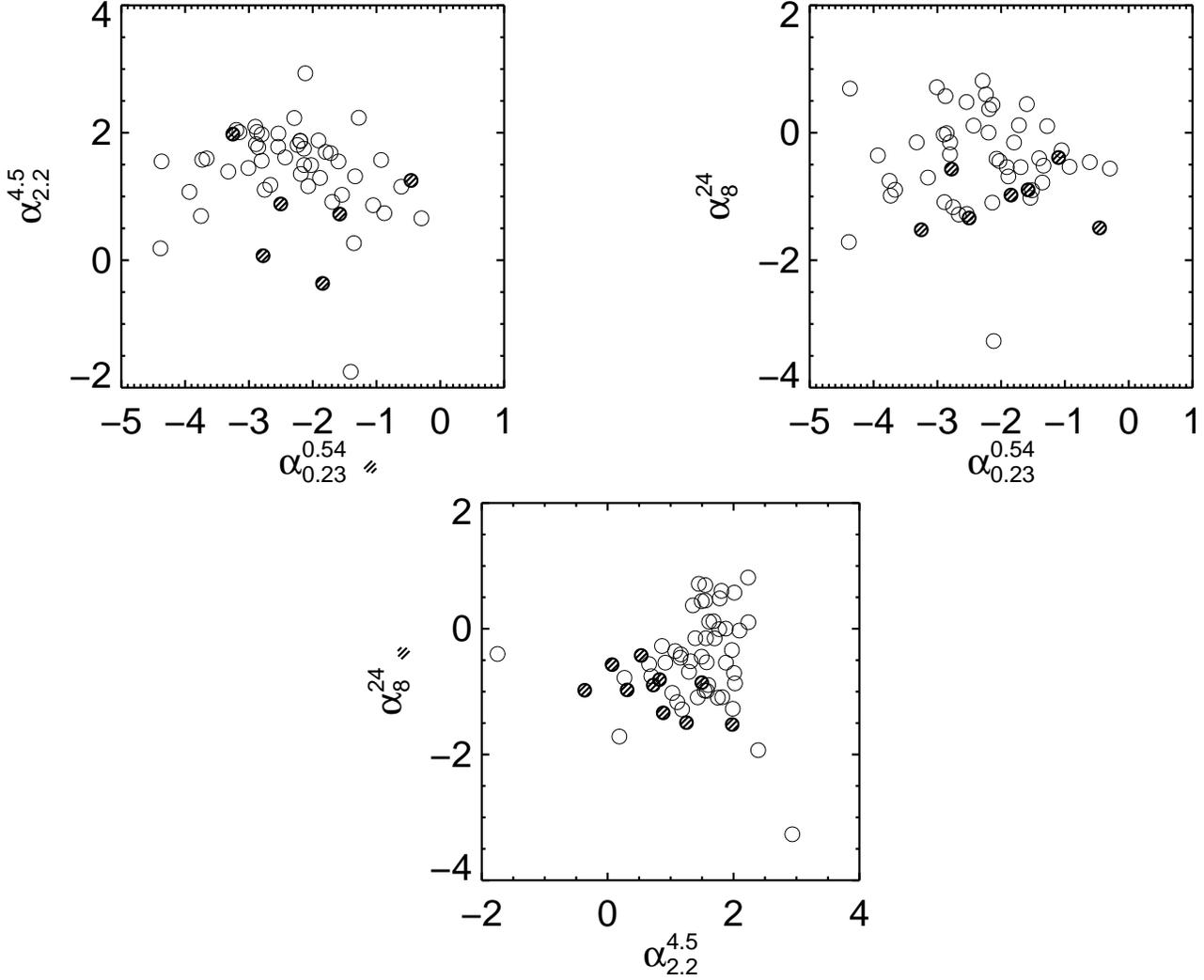}
    \caption{Comparison of  spectral indices ($\alpha_{\rm 0.23}^{0.54}$,
      $\alpha_{\rm 2.2}^{4.5}$, and $\alpha_{\rm 8}^{24}$) for ULIRGs
      (hashed circles) and LIRGs (open circles). In the panels
      with the mid-IR index $\alpha_{\rm 8}^{24}$, the ULIRGs tend to
      cluster in a small region on the plot due to their very dusty
      nature, but the near-IR index $\alpha_{\rm 2.2}^{4.5}$ does not
      work as well in distinguishing the ULIRGs from their
      lower-luminosity counterparts.} 
    \label{fig:alpha_comp}
  \end{figure}

   In \fig \ref{fig:alpha_comp} we compare spectral indices:   
   $\alpha_{0.23}^{0.54}$ vs. $\alpha_{2.2}^{4.5}$
   contrasts  the  slopes on either side of the stellar ``bump";
   $\alpha_{0.23}^{0.54}$ vs. $\alpha_{8}^{24}$ contrasts the blue-ward
   slopes of the optical and the infrared ``bumps", respectively; 
   $\alpha_{2.2}^{4.5}$ vs. $\alpha_{8}^{24}$ contrasts  the red-ward
   and blue-ward slopes of the optical and infrared  ``bumps",
   respectively.  The correlation coefficients for these three sets of
   comparisons are -0.08, -0.03, and -0.01.  With this sample size,
   the conservative, non-directional $p$-values are 0.57, 0.83, and
   0.94, respectively. This indicates that there is no correlation
   found between each pair of spectral indices.  However, we note that
   for all three spectral index comparisons, the ULIRGs tend to show
   smaller values of $\alpha_8^{24}$ and $\alpha_{2.2}^{4.5}$,
   corresponding to a deeper trough at $\lambda \sim 4-8\mu$m,
   presumably due to greater silicate dust absorption of the continuum
   in ULIRGs. The physical significance of using $\alpha_{2.2}^{4.5}$ as an
   AGN indicator is further discussed in \S \ref{agn}.

   \subsection{Flux Ratios}
   \label{Ratios}

\begin{figure}
    \centering
    \includegraphics[width=0.8\textwidth,angle=90]{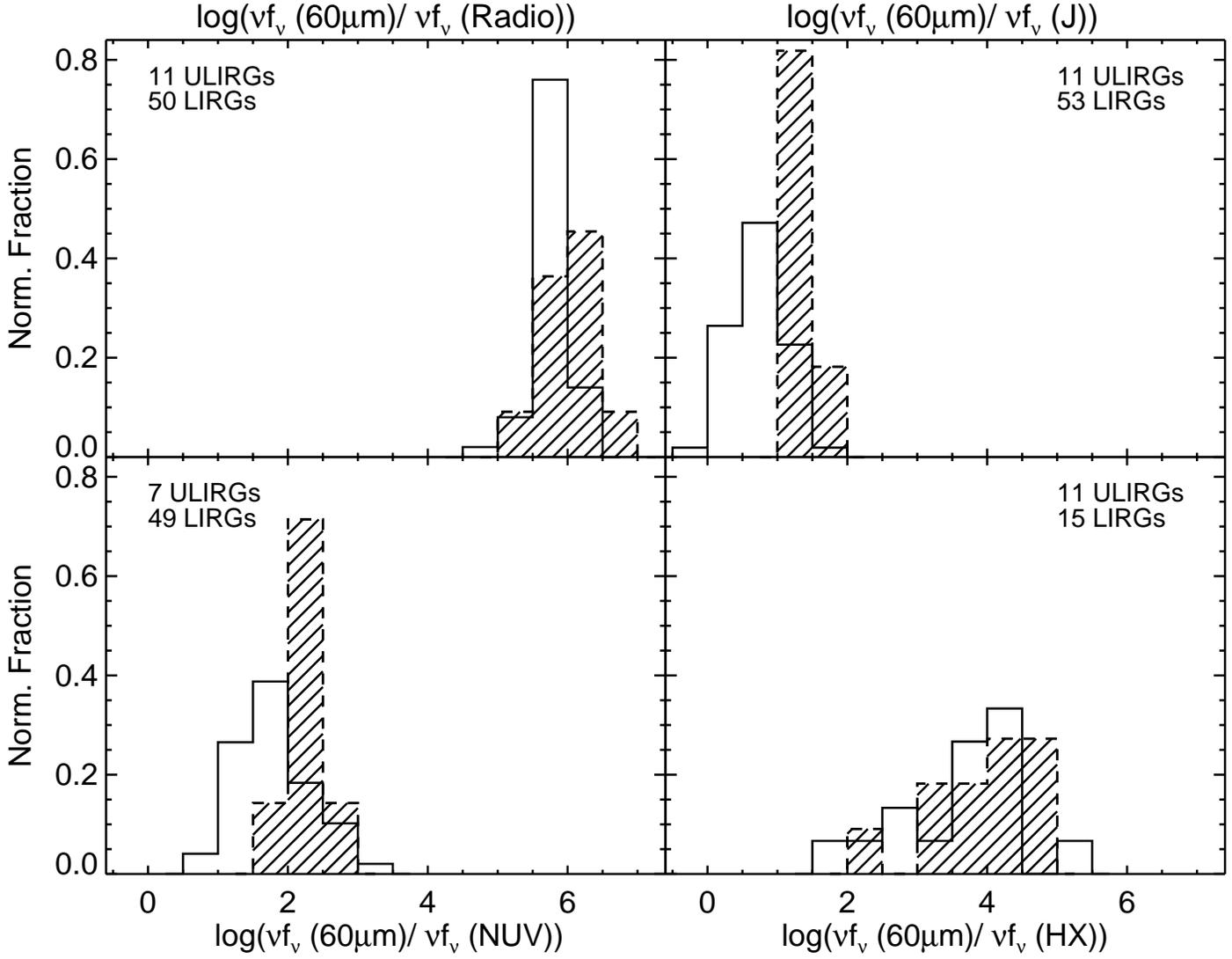}
    \caption{Normalized histograms of logarithmic ratios for $\nu
      f_{\nu}$ at 60$\mu$m to that at radio (1.4 GHz), $J$-band
      (1.2$\mu$m) , NUV (0.23$\mu$m) , and HX (2-10 keV). These
      histograms are normalized by the number of LIRGs (open
      histograms) and ULIRGs (hatched histograms) in
      each plot (as listed in the upper corners), as restricted
      by the availability of the photometry data used in these
      ratios.}
    \label{fig:firratio_lum}
  \end{figure}

   The SED shapes can also be characterized in terms of flux ratios
   with respect to the measured flux at 60$\mu$m to compare direct
   stellar emission to dust emission.  \fig
   \ref{fig:firratio_lum} shows the distribution of flux ratios at
   Radio (1.4 GHz), $J$-band (1.2$\mu$m) , $NUV$ (0.23$\mu$m), and HX
   (2-10 keV), for all 11 ULIRGs and 53 LIRGs, where data are
   available.  The wavelengths chosen for display represent the short
   and long wavelength extremes of the SEDs as well as the ``peak" and
   the short wavelength side of the stellar thermal bump.   For the
   $J$-band and NUV ratios, the difference in the distributions between
   ULIRGs and LIRGs is simply due to the well-known property of the
   SEDs where the thermal stellar ``bump" remains relatively constant
   ($\lesssim 2$) while the thermal dust ``bump" grows by a factor of
   $\sim 10$~\cite[\eg][]{Sanders96}.  The differerent distributions
   for the LIRGs and ULIRGs point to the discrepant starlight-to-dust
   ratios in the two populations. The HX band shows a somewhat
   surprising result in that the ratio seems to be similar for both
   LIRGs and ULIRGs.  

   Finally, we summarize the SEDs of our complete sample of (U)LIRGs
   in \tbl \ref{tbl:stat}, which lists the relative mean luminosity,
   $R= \log [L_\nu(band) / L_\nu(60\mu{\rm m})]$, in all 26 observed
   bands with respect to the mean luminosity at 60$\mu$m.  \tbl
   \ref{tbl:stat} also lists the dispersion in the luminosity ratio,
   $\sigma_R$, and the full range, $\Delta R$. The largest values of
   $R$ are found at the long and short wavelength ends of the SEDs,
   where the emission is clearly not fit by the two (stellar and dust)
   thermal ``bumps".  However, it is interesting to note that a
   small dispersion in $R$ is found in the Radio (1.4 GHz), where
   the ratio is largest ($R= -5.81$).  This would seem to
   confirm that the well-known ``radio-infrared correlation"
   \cite[e.g.][]{Helou85}, also holds for (U)LIRGs.  

   At X-ray wavelengths, there seems to be a 0.4 dex increase in the mean
   luminosity in the HX band compared to the SX band. Of the 26
   objects within our sample that are detected in both X-ray bands,
   $\sim$ 58\% are more luminous in HX than in SX, suggesting the
   presence of a HX ionizing source.  Due to the relative
   incompleteness of X-ray observations at the lower-luminosity end of
   our sample, this is slightly higher than the
   conservative estimate of 37\% AGN fraction (or 48\% if [Ne {\sc v}]
   detection is taken into account) from an X-ray study of 44 GOALS
   (U)LIRGs at the high end of the $L_{\rm IR}$ range~\cite[with
   median $\log (L_{\rm IR}/L_\odot) =
   11.99$;][]{Iwasawa11}. Incorporating the entire 
   GOALS sample,~\cite{Petric11} found that 18\% of all (U)LIRGs contain
   an AGN based on a mid-IR \emph{Spitzer}-IRS study; the
   comparatively lower fraction reflect that fewer of the
   lower-luminosity objects feature an AGN. 
   We note that because of the complexity of dust geometry within these
   systems, discrepancies among the AGN fractions thus determined may
   be due to the limited sensitivity of the various AGN indicators. 

  \section{Physical Properties of the GOALS Sample}
  \label{Discussion}
  Here we discuss the template fitting done to compute infrared
  luminosity and stellar mass; specifically, we fit the 
  MIR-submillimeter portion of the SED with various dust models to
  compute $L[8-1000\mu{\rm 
    m}]$ and compare with previous estimates of $L_{\rm IR}$ computed
  from IRAS photometry.  We use population synthesis models to
  fit the UV-NIR portion of the SED in order to determine stellar
  mass.  These masses are then compared with stellar mass estimates
  computed using $H$-band luminosities alone.  

  \subsection{Template Fits}
   \label{Lephare}
   We have fitted each of the (U)LIRG SEDs with stellar population synthesis
   and dust models (see \fig \ref{fig:allseds}). Our goals are 
   two-fold:\  (1) to better determine stellar mass ($M_\star$) and
   subsequently star formation rate (SFR) of the local (U)LIRGs, and (2)
   to better estimate the flux at any unobserved wavelength band.
   The optical-through-NIR SED fitting has been done using the Le
   PHotometric Analysis for Redshift Estimations (Le
   PHARE\footnote{\url{http://www.cfht.hawaii.edu/$\sim$arnouts/LEPHARE/cfht\_lephare/lephare.html}})
   code, a photometric redshift and simulation package developed by
   S. Arnouts \& O. Ilbert.  It is capable of providing optical and
   FIR fitting as well as a complete treatment of physical parameters
   and uncertainties based on the simple stellar population synthesis
   model of choice.   As described below we consider both a
   Salpeter~\cite[][]{Salpeter55} and Chabrier~\cite[][]{Chabrier03}
   IMF for all of our sources, plus a Calzetti extinction
   law~\cite[][]{Calzetti94} adopted throughout.   Stellar masses were
   determined from fitting the observed data shortward of $K$-band.

   \subsubsection{Infrared Luminosities, Dust Temperatures, and Dust Masses}
   \label{FIRlum}

   The infrared luminosity, $L_{\rm IR}$, that is discussed throughout
   this paper refers to the luminosity emitted in the wavelength range
   $8-1000\mu$m.  The values of $L_{\rm IR}$ given in \tbl
   \ref{tbl:basicinfo} have been adopted from the RBGS using the
   following prescription reproduced from \cite{Perault87} and
   \cite{Sanders96}: 
   \begin{equation}
     F_{\rm IR} = 1.8 \times 10^{-14} (13.45 f_{12} + 5.16 f_{25} +
     2.58 f_{60} + f_{100}) ~\quad\rm{[W m^{-2}]}~,
     \label{eqn:perault}
   \end{equation}
   \begin{equation}
     L[8-1000 \mu m] = 4 \pi D^2_L F_{\rm IR} ~\quad[L_\odot]~,
   \end{equation}
   where $f_{12}$, $f_{25}$, $f_{60}$, $f_{100}$, are the flux
   densities in Jy at 12, 25, 60, and 100 $\mu$m, respectively, and
   $D_L$ is the luminosity distance.

   Given the availability of our new IRAC, MIPS and SCUBA data points,
   we can now test the above \emph{IRAS} approximation and the
   validity of the assumed SED, in particular longward of 100$\mu$m,
   for the local (U)LIRG population.  We compute a new total infrared
   luminosity,  $L_{\rm IR} = 8-1000\mu$m, by using $\chi^2$
   minimization to fit the MIR-FIR-submm portion (14 data points) of
   the SEDs with different dust models and 
   a modified black body fit \cite[CE,DH,SK,mBB:][\tbl
   \ref{tbl:dust}]{Chary01,Dale02,Siebenmorgen07,Casey12}. For the CE
   models, 105 templates with infrared luminosity ranging from $L_{\rm
     IR} = 10^{8.44}$ to $L_{\rm IR} = 10^{13.55}$ have been used. For
   the DH models, the 64 templates employed exhibit infrared
   luminosity within the range of $10^{8.32}-10^{14.34}$. The SK
   template library in LEPHARE consists of spatially integrated SEDs
   computed for starbursts of different radii, total luminosity,
   visual extinction, dust density within hot spots, and the
   luminosity ratio of hot spots to 
   total as a secondary parameter based on~\cite{Siebenmorgen07}. We
   utilized the models with radii = 1kpc, $L_{\rm IR} =
   10^{10.1}-10^{14.7}$, $A_V = 2.2-119$ mag, $n_{hs} = 10^2-10^4$ 
   cm$^{-3}$, and OB luminosity to total luminosity of
   40\%-90\%. Lastly, the model-independent mBB fit is essentially the
   sum of a mid-infrared powerlaw at $\lambda < 50\mu m$ and a
   single-temperature greybody at $\lambda > 50 \mu$m~\cite[][]{Casey12}: 
   \begin{equation}
     S(\lambda) =
     N_{bb}\frac{(1-e^{-(\frac{\lambda_0}{\lambda})^\beta})(\frac{c}{\lambda})^3}{e^{hc/\lambda
       kT}-1} + N_{pl}\lambda^\alpha e^{-(\frac{\lambda}{\lambda_c})^2}
   \end{equation}
   where $T$ is the far infrared dust temperature, $\lambda_0$ is the
   wavelength corresponding to an optical depth of unity, $\lambda_c$
   is the wavelength where the mid-infrared powerlaw turns over,
   $\beta$ is the emissivity, and $\alpha$ is the spectral index of
   the powerlaw component. We let $\beta$ and $\alpha$ vary to
   fit the SED where adequate data exist~\cite[see more details
   in][]{Casey12}. 
   
\begin{figure}
    \centering
    \includegraphics[width=0.8\textwidth,angle=90,bb=0 350 504 684]{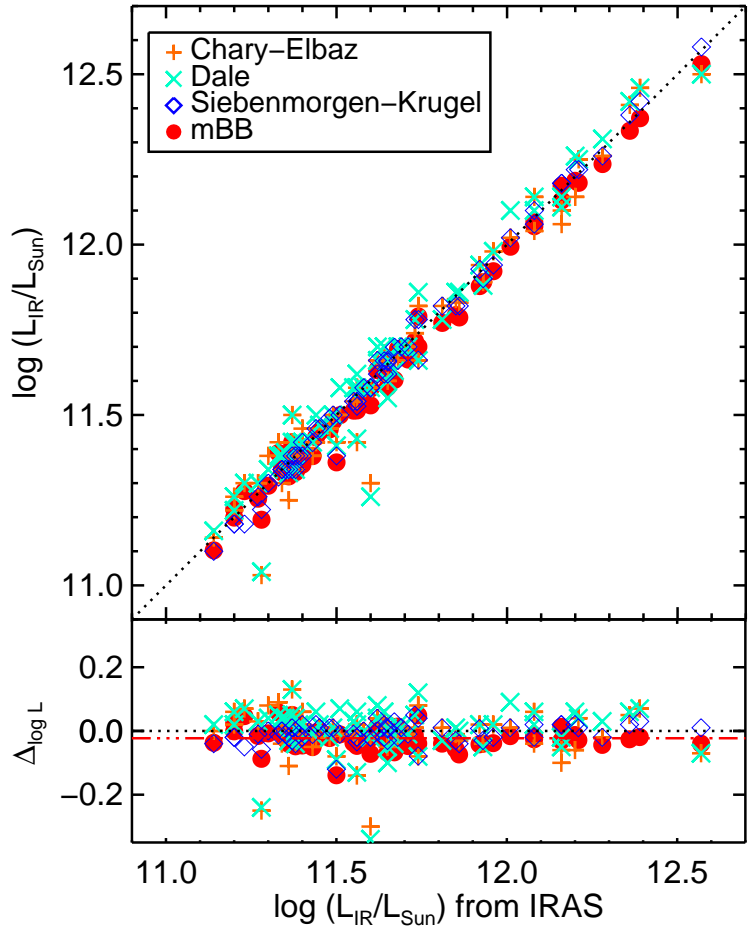}
    \caption{Comparison of the $L_{\rm IR}$ derived from \emph{IRAS}
      fluxes and from fitting the FIR--submillimeter part of the SEDs
      for the (U)LIRGs (CE: orange cross, DH: cyan X, SK: blue open
      diamonds, and mBB: red filled circles).  The dotted lines
      indicate linear correlation with zero offsets. As a one-to-one
      correspondence is expected, the residual $\Delta_{\log L} \equiv
      \log (L_{\rm fit} / L_\odot) - \log (L_{\rm IRAS} / L_\odot)$
      shows the scatter around the values predicted from
      \emph{IRAS}. The red dash-dot line shows that the mean of the
      mBB values is offset by 0.021 dex.} 
    \label{fig:lumcomp}
  \end{figure}

   The comparison of $L_{\rm IR}$ values is shown in \fig
   \ref{fig:lumcomp}. The CE and DH model fits have large scatters
   (0.07 dex and 0.08 dex) around the \emph{IRAS} values. The SK model
   fits exhibit a large scatter (0.05 dex) around the systematic
   offset of $-0.03$ dex from the \emph{IRAS} luminosities.  The mBB
   fit values are $\sim$0.02 dex systematically lower than the
   \emph{IRAS} values, which translates to $\sim$5\% of the infrared
   luminosity at $L_{\rm IR} = 10^{12} L_\odot$ level.  
   The disagreement with the \emph{IRAS} values is primarily due to
   the inadequate color assumptions implied in the coefficients of
   \eqn \ref{eqn:perault}. The scatter and discrepancy exhibited by the
   luminosities derived from the model-dependent templates are due to
   limitations in the step size within the model grids. 

   Comparing the
   fits of the different templates and the mBB at 12, 25, 60, 100, and
   850$\mu$m, the residuals between the data and the different fits
   are the smallest for mBB. The SK fits result in similarly minimal
   residuals except at the short wavelengths, whereas the CE and DH
   fit residuals exhibit the largest scatters. The main reason for the
   mismatch to the data is that the detailed infrared SED models have many
   degrees of freedom and very large template libraries at discrete
   temperatures and other grid parameters. In contrast, the analytical mBB
   fit provides more fitting flexibility and is designed to represent
   the data as accurately as possible, despite having only three free
   parameters and not accounting for the mid-infrared PAH
   features. Since for this paper we only measure $L_{\rm IR}$,
   $T_{\rm dust}$, and $M_{\rm dust}$, we choose to adopt the mBB fit
   results for the rest of the analyses.  

   Fitting greybody of a single dust temperature to
   the FIR SED from the blackbody peak ($\sim 100\mu$m) to
   1000$\mu$m, we determine and present the temperature of the dust in
   \tbl \ref{tbl:dust}: mean $T_{\rm dust} = 33.2\pm6.2$ K. This
   temperature is $\sim$10K cooler than that determined from
   the~\cite{Perault87} prescription~\cite[see discussion
   in][]{Casey12}.  With the lack of long-wavelength data points, the
   latter assumes a single temperature dust emissivity model
   $\epsilon~\propto~\nu^{-1}$ fit to the fluxes in the four \emph{IRAS}
   bands~\cite[as described in][]{Sanders96}, often adopting a peak
   shortward of the true peak now revealed when data points longward 
   of 100$\mu$m are present.  The MIPS-160$\mu$m and the
   SCUBA-850$\mu$m points are invaluable for constraining the real peak,
   and this will be nailed when the \emph{Herschel} far-infrared
   imaging observations are completed for the entire GOALS sample.
   Dust masses were calculated using the 850$\mu$m flux from the SED
   and \eqn 8 in~\cite{Casey11}:
   \begin{equation}
     M_{\rm dust} = \frac{S_\nu~D^2_{\rm L}}{\kappa_\nu~B_\nu(T)}
     \propto \frac{D^2_{\rm L}}{\kappa_\nu~\nu^2} \quad ,
   \end{equation}
   where $S_\nu$ is the flux density at frequency $\nu$, $\kappa_\nu$
   is the dust mass absorption coefficient at $\nu$, $B_\nu(T)$ is the
   Planck function at temperature $T$, and $D_{\rm L}$ is the
   luminosity distance.  We adopted a dust absorption coefficient of
   $\kappa_{850} = 0.15$ m$^2$
   kg$^{-1}$~\cite[][]{Weingartner01,Dunne03}. We derived a mean dust
   temperature $M_{\rm dust} = 10^{7.3} M_\odot$ for the sample.   

We note that dust temperatures and
   masses depend critically on a thorough understanding of the radiative
   transfer involved as well as the underlying geometry of the dust cloud.
   This dust temperature can be taken as a characteristic $T_{\rm
     dust}$ for the system (measured from the peak of the SED).
   Clearly the physics of these galaxies is more complex, comprising
   of many dust reservoirs of different temperatures. Unfortunately
   probing that is beyond the current scope of observations and this work.
   We included our derivation of the dust temperature primarily as a
   baseline for comparison to the high-$z$ universe, though we caution
   that the physical interpretation of that quantity remains uncertain.

   \subsubsection{Stellar Masses}
   \label{Masses}

   Stellar masses for the (U)LIRGs in our sample have been computed via
   two methods using two different initial mass functions (IMFs),
   and the resulting masses are given in \tbl \ref{tbl:masses}.  Here,
   we discuss the differences among these measurements. 

   The two methods adopted for mass determination were to fit the
   UV-NIR part of the SEDs and to scale from the $H$-band
   luminosity. $H$-band is usually selected for stellar mass
   conversion because it is at or near the photospheric peak of the
   stellar SED, and is thought not to be contaminated by hot dust
   emission from AGN~\cite[][]{Hainline11}.  However, problems with
   $H$-band scaling may also arise 
   due to thermally pulsing asymptotic giant branch stars for SEDs
   with a significant contribution from young stellar
   populations~\cite[][]{Walcher11}. On the other hand, the
   SED-fitted masses encompass the stellar component contributing to
   the optical peak, taking into account the treatment of dust with
   the designated dust extinction law. For the ensuing discussion, we
   focus on comparing the SED-fitted masses derived from using two
   different IMFs. 

\begin{figure}
    \centering
    \includegraphics[width=0.8\textwidth,angle=90,bb=0 350 504 684]{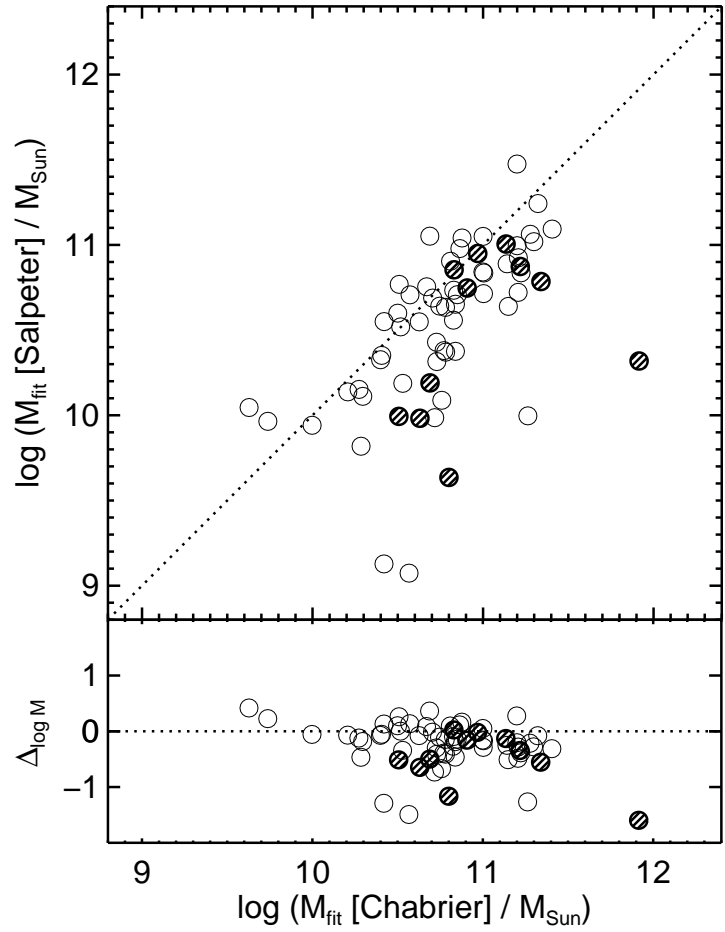}
    \caption{Comparison of the stellar mass estimates derived from fitting the
      UV-NIR part of the Salpeter-based BC03 SEDs and from that of the
      Chabrier-based ones.  The residual plot illustrates that $\Delta_{\rm
        \log M} \equiv \log (M_{\rm Sal} / M_\odot) - \log (M_{\rm Chab} /
      M_\odot)$ centers at -0.26, with the 25$^{th}$ percentile at
      -0.41 and the 75$^{th}$ percentile at -0.02. The open circles are
      LIRGs and hatched circles are ULIRGs.  The dotted lines
      indicate linear correlation with zero offset.}
    \label{fig:masscomp_offset}
  \end{figure}

   Our optical-NIR SED fitting procedure is based on stellar
   population synthesis models from \cite{Bruzual03} using 10 different
   broadband UV, optical, and infrared bands.  Two different IMFs were
   used: a Chabrier IMF~\cite[][]{Chabrier03} and a Salpeter
   IMF~\cite[][]{Salpeter55}. Since the IMF dictates the scaling of
   the mass-to-light ($M/L$) ratio in converting luminosity to mass via
   its slope and mass cutoffs, we compare the masses derived from
   these two IMFs and assess their differences (\fig
   \ref{fig:masscomp_offset}).  
   The lower and upper mass cutoffs employed were 0.1 $M_\odot$ and
   100 $M_\odot$, respectively~\cite[][]{Bruzual03}; no additional
   adjustments to the parameters of the IMFs have been made.
   The models assumed a star formation history with $SFH =
   e^{-t/\tau}$ where $\tau$, varying from $1-30$, is the e-folding
   parameter in years.  The metallicity ($Z = 0.004, 0.008, 0.02$) has
   been treated as a free parameter as well.

   In general, the mean difference between the masses derived from
   both methods is at the level of 0.26$\pm$0.41 dex, with the masses
   generally being underestimated from the Salpeter IMF. 
   This was unexpected given that Salpeter IMF tends to result in
   higher stellar masses due to the difference in treatment of the
   low-mass end --- the Chabrier IMF tends flatter and therefore more
   physical. Under the same input parameters the stellar
   models with Chabrier IMF fitted the UV light in our SEDs better
   than their Salpeter counterparts. 
   We also note that the Chabrier IMF incorporates a more up-to-date
   treatment of UV radiation from young stars in starburst
   populations, and thus the Chabrier masses have been adopted as the
   stellar masses for the local (U)LIRGs.  

 \subsubsection{Comparison of $M_\star$ and $L_{\rm IR}$}
   \label{CompML}

\begin{figure}
    \centering
    \includegraphics[width=0.8\textwidth,angle=90]{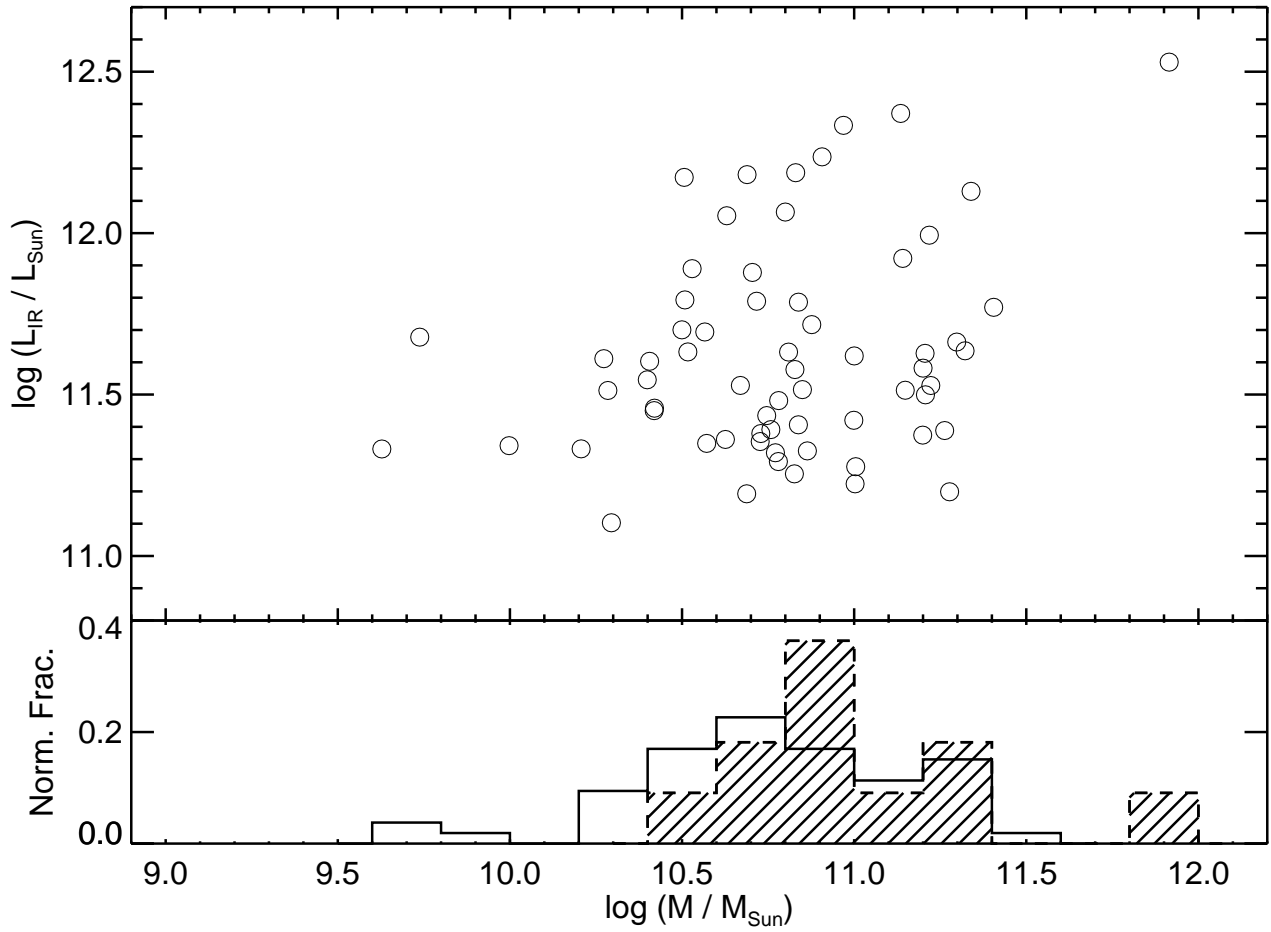}
    \caption{The mBB infrared luminosity vs. stellar mass as derived
      from template fitting in logarithmic scale for our sample. The
      range of the masses spans more than a factor of 10 for the
      LIRGs. This is more clearly visualized in the histogram at the
      bottom, with bin width of 0.15 dex. The open histogram with
      solid line represents the LIRGs, while the hatched histogram
      represents the ULIRGs.}  
    \label{fig:masslum}
  \end{figure}

   \fig \ref{fig:masslum} shows the (U)LIRGs as a distribution of
   SED-fitted mass in logarithmic scale.  All but three of the 64
   objects fall within 2$\sigma$ of the mean mass, log
   $(M_\star/M_\odot) = 10.79\pm0.40$.  
   
   For the individual subsamples of 53 LIRGs and 11 ULIRGs, we find
   the mean stellar masses to be 
   $\log (M_\star/M_\odot) = 10.75 \pm 0.39$ for the LIRGs, and 
   $\log (M_\star/M_\odot) = 11.00 \pm 0.40$ for the ULIRGs.
   The factor of $\sim 2$ lower mean stellar mass for LIRGs is 
   primarily due to a decreasing ``low mass tail" of objects with
   masses in the range $\log (M_\star/M_\odot) = 9.3 - 10.3$.  On the
   other extreme, the most massive object (UGC 08058 = Mrk 231)
   appears to show hot dust emission from the central AGN, which
   contributes about 20\% of the $H$-band
   flux~\cite[][]{Surace99,Surace00b}.  Thus for this one object, we
   have corrected the fitted stellar mass accordingly.
   
   Our results are consistent with estimates of the mass of
   objects of similar luminosity at higher redshift: log~$(M_{\rm
     ULIRG}/M_\odot)\,\sim\, 10.51 \pm 0.45$~\cite[from][where the
   two-tailed unpaired $p$-value for the differences between our ULIRG
   masses gives 0.052, which is not statistically
   significant]{Takagi03}; and  log~$(M_{\rm LIRG}/M_\odot)\,\sim\,10.5$
   ~\cite[from][who found that (U)LIRGs are more massive than
   ``normal'', non-LIRG galaxies that are morphologically irregulars
   and spirals from the GOODS-S field 
   where log~$(M_{\rm normal}/M_\odot)\,\sim\,9.5$]{Melbourne08}.

   \subsubsection{Star Formation Rates}
   \label{SFR}

\begin{figure}
    \centering
    \includegraphics[width=0.8\textwidth,angle=90]{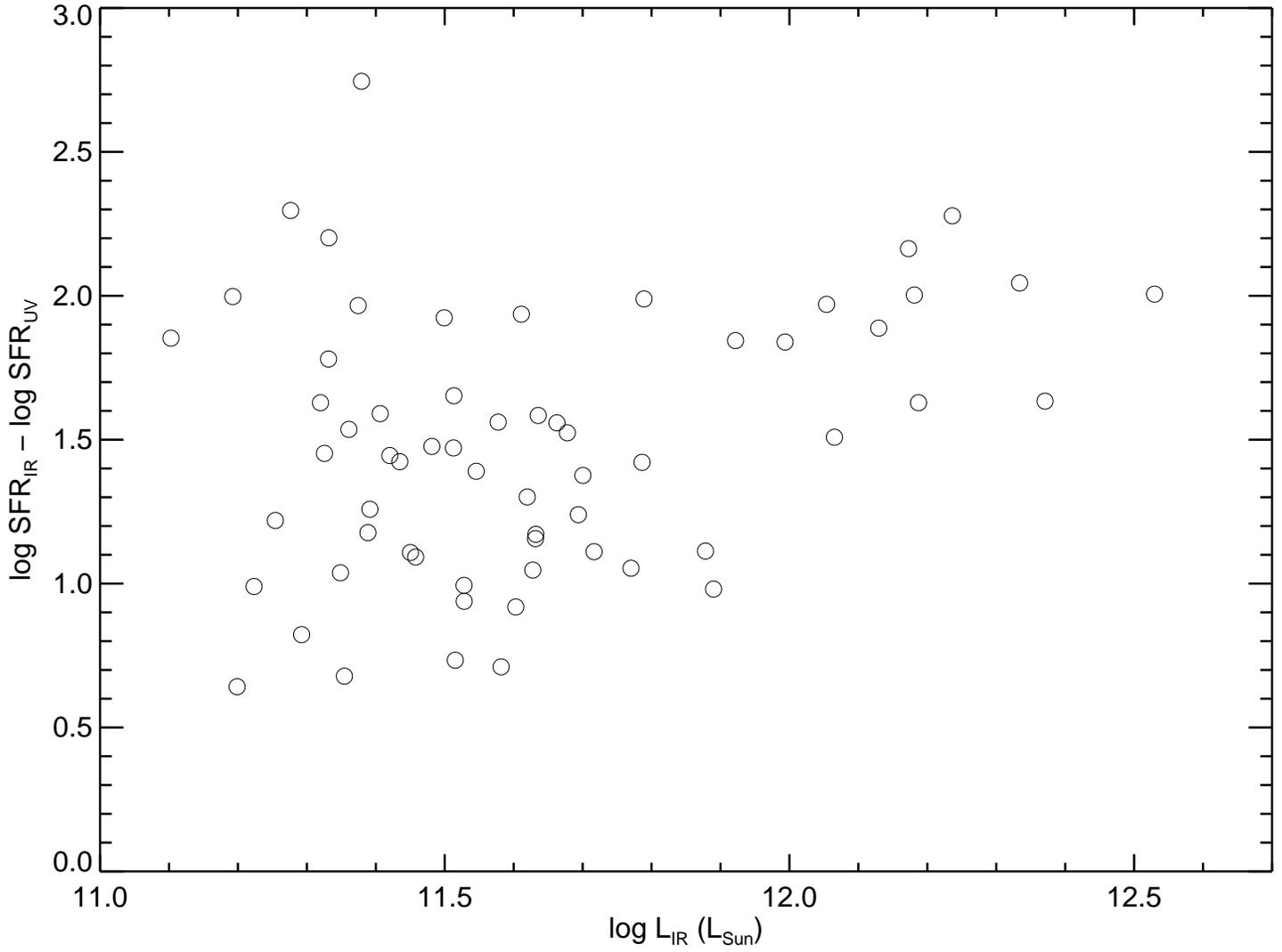}
    \caption{The logarithmic difference in SFR determined from the IR
      and the UV vs. $L_{\rm IR}$. The positive values in this
      difference reflect that the UV-derived SFR underpredict the SFR
      as measured from IR light.} 
    \label{fig:sfr}
  \end{figure}

   Using the light contribution from the UV and IR, we determine the
   star formation rate for the unobscured and obscured stellar
   populations, respectively. The recipe from~\cite{Wuyts11} gives the
   following calibration for converting from infrared and
   monochromatic UV luminosity at 2800\AA~to star formation rate:  
   \begin{equation}
     {\rm SFR_{UV+IR}} [M_\odot {\rm yr}^{-1}] = 1.09 \times 10^{-10}
     (L_{\rm IR} + 3.3 L_{2800})/L_\odot \quad. 
   \end{equation}
   Decomposing this quantity into separate UV and IR components, we
   consider the contribution to the total star formation rate from the
   UV and IR luminosities individually (\tbl \ref{tbl:sfr}). SFR$_{\rm
     UV}$ ranges from $<$1 to $\sim$10 $M_\odot$ yr$^{-1}$, while
   SFR$_{\rm IR}$ is up to $\sim$50 times larger.  We show the fold
   enrichment of SFR$_{\rm IR}$ to SFR$_{\rm UV}$ as a function of
   infrared luminosity in \fig \ref{fig:sfr}. The logarithmic
   difference for the LIRGs centered at 1.49$\pm$0.66 dex, while 
   that for the ULIRGs is more clustered at 1.90$\pm$0.24 dex. This
   figure highlights that while the infrared star formation rate in ULIRGs
   is $\sim$100 times that determined from the UV, the fold enrichment
   in the LIRGs is only $\sim$30 times with a large scatter
   potentially due to the large variations in dust geometry (\ie
   single spirals undergoing minor merger events as opposed to
   major-merging pairs). 

   The mean log SFR$_{\rm UV+IR}$ for the GOALS ULIRGs is
   2.25$\pm$0.16 and that for the LIRGs is
   1.57$\pm$0.19. Corresponding specific star formation rates (sSFR = 
   SFR$_{\rm UV+IR}$/$M_\star$) have been computed and are also listed
   in \tbl \ref{tbl:sfr}.  
   The effect on SFR from UV emission 
   and trends with infrared luminosity as seen in the GOALS (U)LIRGs
   as well as the comparison to a nearby lower-luminosity
   sample~\cite[SINGS;][]{Kennicutt03} have been
   discussed extensively in~\cite{Howell10}. 
   Compared with the~\cite{Howell10} sample (with median SSFR $= 3.9
   \times 10^{-10}$ yr$^{-1}$), the median SSFR of our sample is $6.8
   \times 10^{-10}$ yr$^{-1}$, or equivalent to a mass-doubling
   timescale of 1.5 Gyr. Our median SSFR value is higher because the
   \emph{GALEX} sample is more complete at the lower-luminosity end
   (with less extinction by dust).  The slope of the $\log
   {\rm SSFR} - \log M_\star$ relation is -0.78 (or 0.22 in
   $\log {\rm SFR} - \log M_\star$ space) for the (U)LIRGs,
   which is shallower than that reported for the high-$z$ infrared
   main sequence galaxies~\cite[][]{Rodighiero11,Daddi07}.  

   \subsubsection{AGN Indicators}
   \label{agn}

\begin{figure}
    \centering
    \includegraphics[width=0.95\textwidth]{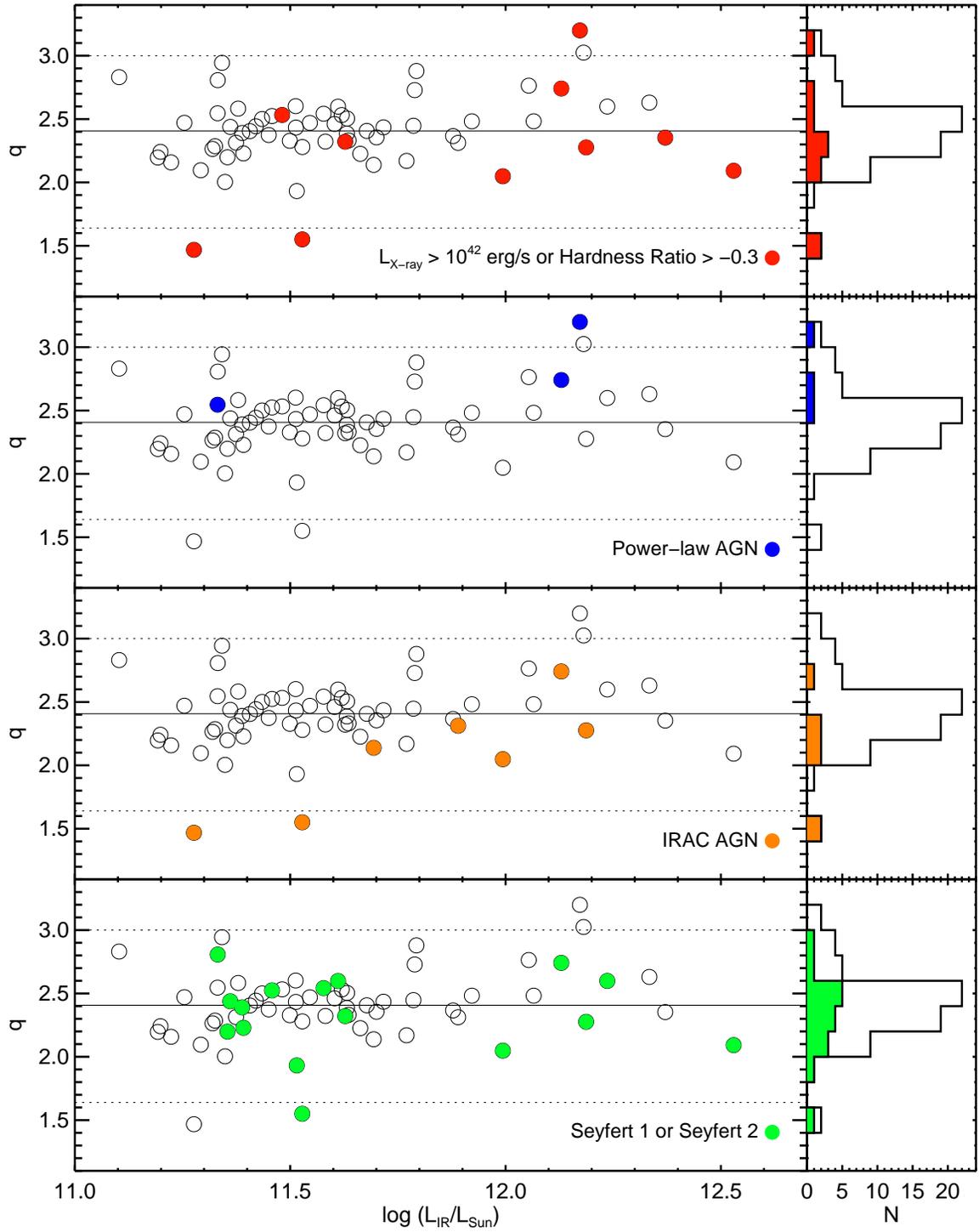}
    \caption{$q$ vs. $\nu L_{\rm IR}$ plots shown for the 64 (U)LIRGs
      along with the distribution in $q$ on the right panels. The
      solid line indicates the median value ($q$ = 2.41) of the sample,
      with $\sigma$ = 0.29. The dotted lines indicate the radio-excess
      ($q < 1.64$) and infrared-excess ($q > 3.0$) criteria
      from~\cite{Yun01}.  From top to bottom are four different AGN
      indicators (filled circles/histograms): X-ray luminosity and
      hardness ratio, power-law slope, IRAC colors, and optical
      spectral classification.}  
    \label{fig:qlir}
  \end{figure}

   Different wavelengths offer different methods for diagnosing AGN
   candidates; a multi-wavelength SED study allows simultaneous access to
   these various indicators and may be used to evaluate their
   effectiveness.  We employ the radio-infrared flux ratio ($q$) as
   defined by~\cite{Condon91} and the criteria specified
   by~\cite{Yun01} (radio-excess: $q < 1.64$; infrared-excess: $q >
   3.0$) as the basis of our comparison. \fig \ref{fig:qlir} shows 
   $q$ plotted as a function of $L_{\rm IR}$ along with its
   distribution in the right panels. The mean $<q> = 2.41\pm0.29$ for
   the sample.  With these limits, there are two
   LIRGs (with $L_{\rm IR} < 11.6$) identified as radio-excess
   sources.  These objects may be potential AGN hosts with compact
   radio core or radio jets and lobes~\cite[][]{Kartaltepe10,Sanders96}.
   On the other end, both of the infrared-excess sources are
   ULIRGs, which may be hosting dense and compact starbursts, or a
   dust-enshrouded AGN.  

   For objects with X-ray detection, we define two different criteria
   for identifying X-ray AGNs: 1. $L_{\rm HX} > 10^{42}~{\rm erg
    ~s}^{-1}$~\cite[][]{Kartaltepe10}, and 2. Hardness Ratio HR $>
   -0.3$~\cite[][]{Iwasawa11}. Seven systems qualified as an X-ray AGN
   by the first criterion, six by the second, and three of these
   objects were identified by both.  Both radio-excess sources are
   very luminous in the HX band.  All the HR $> -0.3$ AGN
   candidates are ULIRGs, though there may be a selection bias since
   all 11 of 11 ULIRGs in the GOALS sample have been observed and
   detected in the X-ray, but only 16 of 53 LIRGs have been observed
   thus far. \emph{Chandra} observations of the lower-luminosity LIRGs
   have been proposed and awarded in Cycle 14 (P.I. Sanders) to
   complete the sample. 

   Power-law SED provides a complementary way to select AGNs that
   might be heavily obscured and opaque to hard X-ray
   emission. We apply a criterion based on $\log(\nu L_{4.5}) -
   \log(\nu L_{2.2}) > 0$ to select galaxies with power-law SED shape
   in the near-IR, corresponding to $F_\nu$ spectral slope of
   0.4~\cite[][]{Kartaltepe10,Alonso06,Donley07}.  Only three
   power-law AGN candidates are identified in the sample, and the two
   ULIRGs among these are both X-ray AGN candidates.  %The object that
%   is a power-law AGN, an X-ray AGN, and an infrared-excess source is
%   IRAS F08572+3915.  

   Additional MIR-based selection criteria have been devised to
   identify heavily-obscured AGNs missed in deep X-ray surveys.  In
   particular, selection based on IRAC color cuts
   \cite[][]{Lacy04,Stern05} is insensitive to obscuration but can
   trace AGN-heated dust, providing a powerful technique for
   discerning luminous obscured and unobscured AGNs.  The revised IRAC
   color selection by~\cite{Donley12} is designed to incorporate the
   best aspects of the current IRAC wedges but to minimizie
   star-forming contamints, and has been adopted here for selecting
   IRAC AGN candidates~\cite[see \eqn 1 and 2 in][]{Donley12}.  
   This set of criteria results in 7 IRAC AGN candidates in our
   sample, two of them being ULIRGs. Interestingly, most of the IRAC
   AGNs are below the median $q$ of the sample ($q_{\rm med} = 2.41$),
   with two of them being radio-excess sources.  

   Our last AGN indicator comparison involves selection based on
   optical emission line diagnostics.~\cite{Yuan10} applied an
   SDSS-based semi-empirical optical spectral classification scheme to
   a large sample of local infrared galaxies, 57 of which are in our
   current sample. Among these, two are Seyfert 1 galaxies and 14 are
   Seyfert 2 systems.  These Seyferts bear a distribution in $q$
   similar to that of the entire sample.

  \section{Conclusions}
  \label{Summary}

  We have used common aperture masks for the first time to assemble 
  accurate radio through X-ray SEDs of a complete local sample of 53
  LIRGs and 11 ULIRGs observable from the northern hemisphere ($\delta
  > -30^\circ$, $\vert b \vert > 30^\circ$). We have utilized  several
  new large datasets provided by GOALS's space- and ground-based
  observations of (U)LIRGs along with additional archival data from
  the literature and our own previously unpublished ground-based
  opt-NIR data from Mauna Kea in our analysis of these SEDs.  We
  summarize our findings as follows:  

 \begin{itemize}

    \item The SEDs for all objects are similar in that they show a
     broad, thermal stellar peak ($\sim 0.3-2\mu$m) and a dominant
     FIR ($\sim 40-200\mu$m) thermal dust peak, where $\nu f_\nu
     ({\rm 60\mu m}) / \nu f_\nu(V)$ varies from $\sim 2-30$ with
     increasing $L_{\rm IR}$. The logarithmic difference between the
     stellar and dust peaks is 1.2 dex for the ULIRGs and 0.7 dex
     for the LIRGs.
      
    \item When normalized at IRAS-60$\mu$m, the largest range in the
      luminosity ratio, $R (\lambda) \equiv {\rm log} [\frac{\nu L_\nu
        (\lambda)}{\nu L_\nu (60\mu{\rm m})}]$, observed over the full
      sample is seen in the Hard X-rays ($HX = 2-10$keV), where
      $\Delta R_{HX} = 3.73$ ($\bar R_{HX} = -3.10$).  A small
      range is found in the Radio (1.4 GHz), $\Delta R_{\rm 1.4 GHz} =
      1.75$, where the mean ratio is largest, ($\bar R_{\rm 1.4 GHz} = -5.81$).
    
    \item Infrared luminosities, $L_{\rm IR} (8-1000)\mu$m), have been
      recomputed using a modified blackbody fit~\cite[][]{Casey11} to
      the MIR-FIR-submm SEDs.  The new $L_{\rm IR}$ values are overall
      $\sim$0.02 dex lower than the original~\emph{IRAS}
      values~\cite[][]{Sanders03}, primarily due to the disagreement
      with the color indices implied in the coefficients
      of the~\cite{Perault87} equation. The simple, analytical
      blackbody fit results have thus been adopted in determining the
      FIR properties of the (U)LIRGs (\eg $L_{\rm IR}$, $T_{\rm
        dust}$, and $M_{\rm dust}$). 
       
    \item The stellar masses computed using BC03 for LIRGs,
      $\log(M_{\rm LIRGs}/M_\odot) = 10.75 \pm 0.39$, and ULIRGs,
      $\log(M_{\rm ULIRGs}/M_\odot) = 11.00 \pm 0.40$, are consistent
      with mass estimates of higher redshift LIRGs from
      \cite{Melbourne08} and with mass estimates of higher redshift
      ULIRGs from \cite{Takagi03}, respectively.   

    \item Star formation rates determined from infrared and
      monochromatic UV luminosities individually have been compared:
      SFR$_{\rm UV}$ ranges from $<$1 to $\sim$10 $M_\odot$ yr$^{-1}$,
      while SFR$_{\rm IR}$ is up to $\sim$50 times larger. The
      logarithmic difference for the ULIRGs is much more clustered
      (1.90$\pm$0.24 dex) than for the LIRGs (1.49$\pm$0.66 dex),
      plausibly due to large variations in dust geometry among the
      lower-luminosity objects.  

    \item  Radio---infrared flux ratio ($q$), along with other
      multiwavelength criteria, have been assessed as different AGN
      indicators.  The results among the various selection techniques
      complement each other. 
      About 60\% of the ULIRGs (and 25\% of the LIRGs) would be
      classified as an AGN by at least one of the selection criteria.
  
  \end{itemize}

\acknowledgements

\begin{figure}
    \centering
    \includegraphics[width=0.9\textwidth]{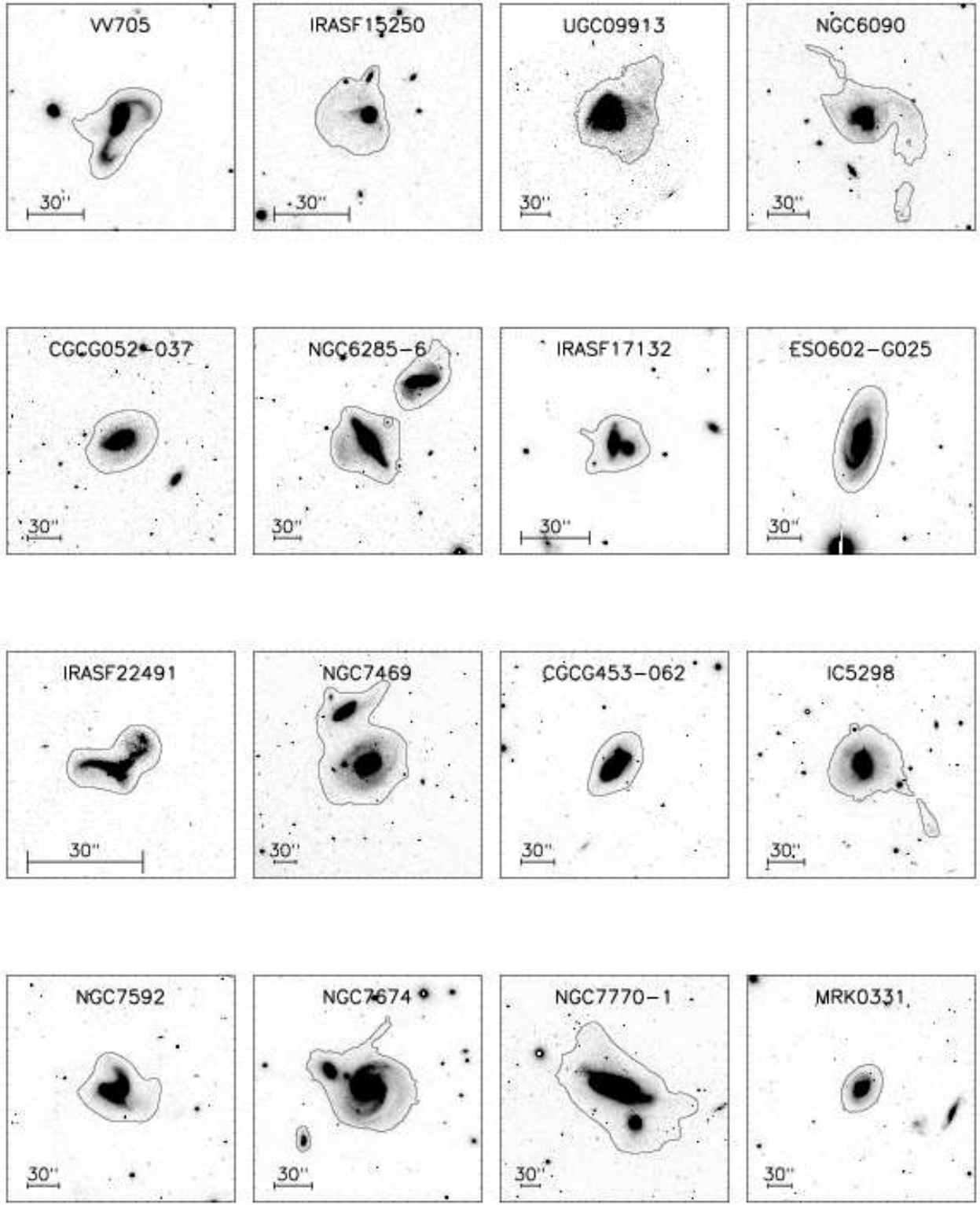}
    \caption{\emph{HST} and UH-2.2m $I$-band images in increasing RA
      order for the 64 local (U)LIRGs with mask photometry contour
      superimposed. The field-of-view for all images is 100kpc
      $\times$ 100kpc, and a 30'' scale bar is drawn inside each frame
      to help guide the eye.}  
  \label{fig:masks}
  \end{figure}

\begin{figure}
    \centering
    \includegraphics[width=0.8\textwidth,angle=90]{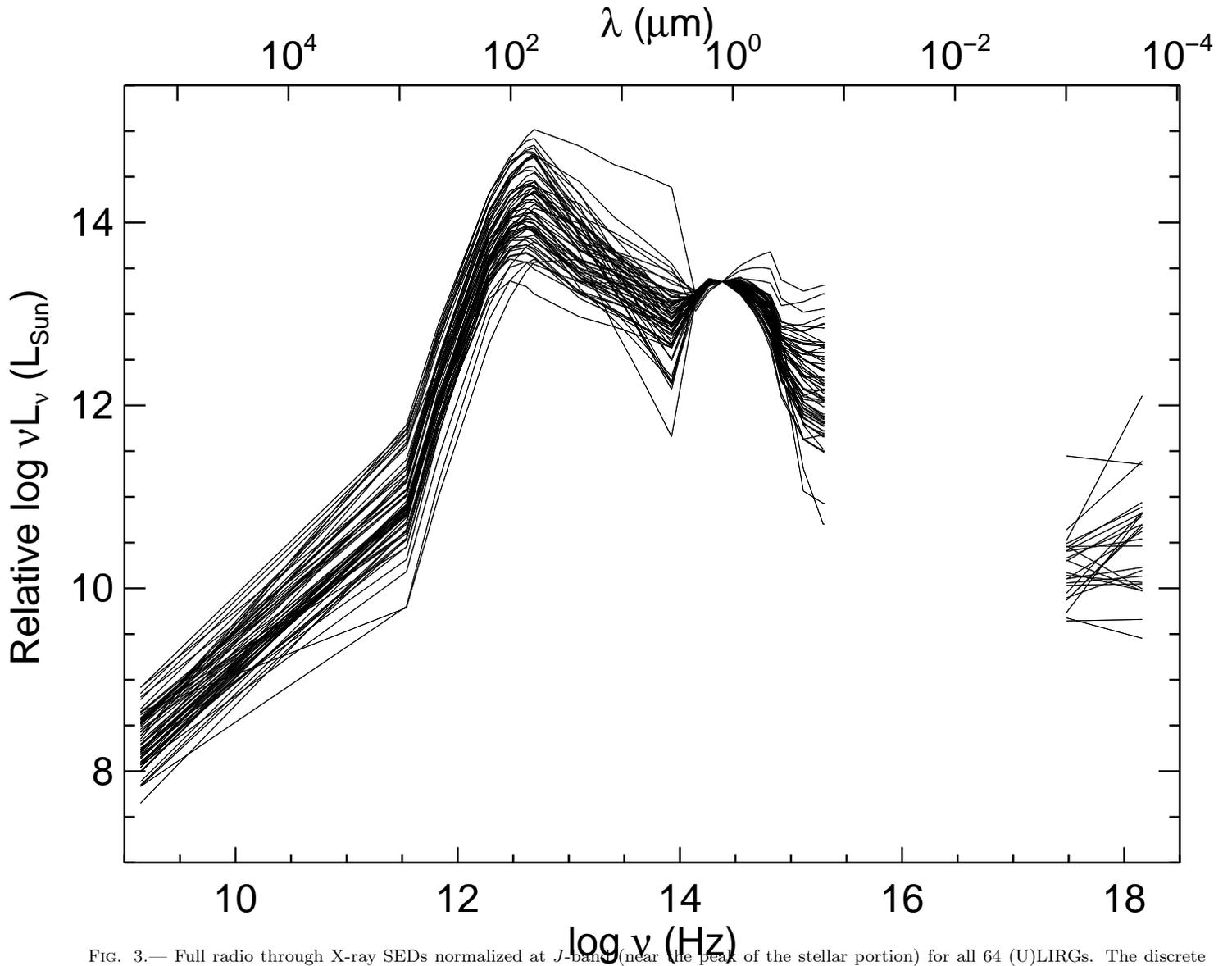}
    \caption{Full radio through X-ray SEDs normalized at $J$-band
      (near the peak of the stellar portion) for all 64
      (U)LIRGs. The discrete data points represent photometry fitted
      at the 27 observed bands. Lines connecting the data points have
      been drawn to help guide the eye except for the regime between
      the far-UV and soft X-ray bands.} 
    \label{fig:seds_stack_nulnu}
  \end{figure}

Vivian U would like to thank O. Ilbert and S. Arnouts for their
help with using the Le PHARE code, C.~J. Ma for his help with UH 2.2m
data acquisition and reduction, T.-T. Yuan for her help with
various scientific and technical contributions, and C.~W.~K. Chiang
for statistical consult and technical help with \fig \ref{fig:masks} \&
\ref{fig:seds_stack_nulnu}. VU also extends 
appreciation towards the UH TAC for their generous support of this
project in awarding telescope time on Mauna Kea, as well as Colin
Aspin and the UH 2.2m Telescope staff for their help and support in
the acquisition of the ground-based optical photometry.  
This research has made use of the NASA/IPAC Extragalactic Database
(NED) and IPAC Infrared Science Achive, which are operated by the Jet
Propulsion Laboratory, California 
Institute of Technology, under contract with the National Aeronautics
and Space Administration. This publication has also made use of data products
from the Two Micron All Sky Survey, which is a joint project of the
University of Massachusetts and the Infrared Processing and Analysis
Center/California Institute of Technology, funded by the National
Aeronautics and Space Administration and the National Science
Foundation. VU wishes to acknowledge funding support from the NASA Harriet
G. Jenkins Predoctoral Fellowship Project and Giovanni Fazio via the
Smithsonian Astrophysical Observatory Predoctoral Fellowship and JPL
Contract/IRAC GTO Grant \# 1256790.  This paper is dedicated to
the memory of Michele Dufault, who put in much hard work into the
photometry extraction for the basis of this paper.

\bibliography{ulirgs}
\bibliographystyle{apj}

\newpage

 \begin{figure}
 \centering
 \includegraphics[width=0.9\textwidth]{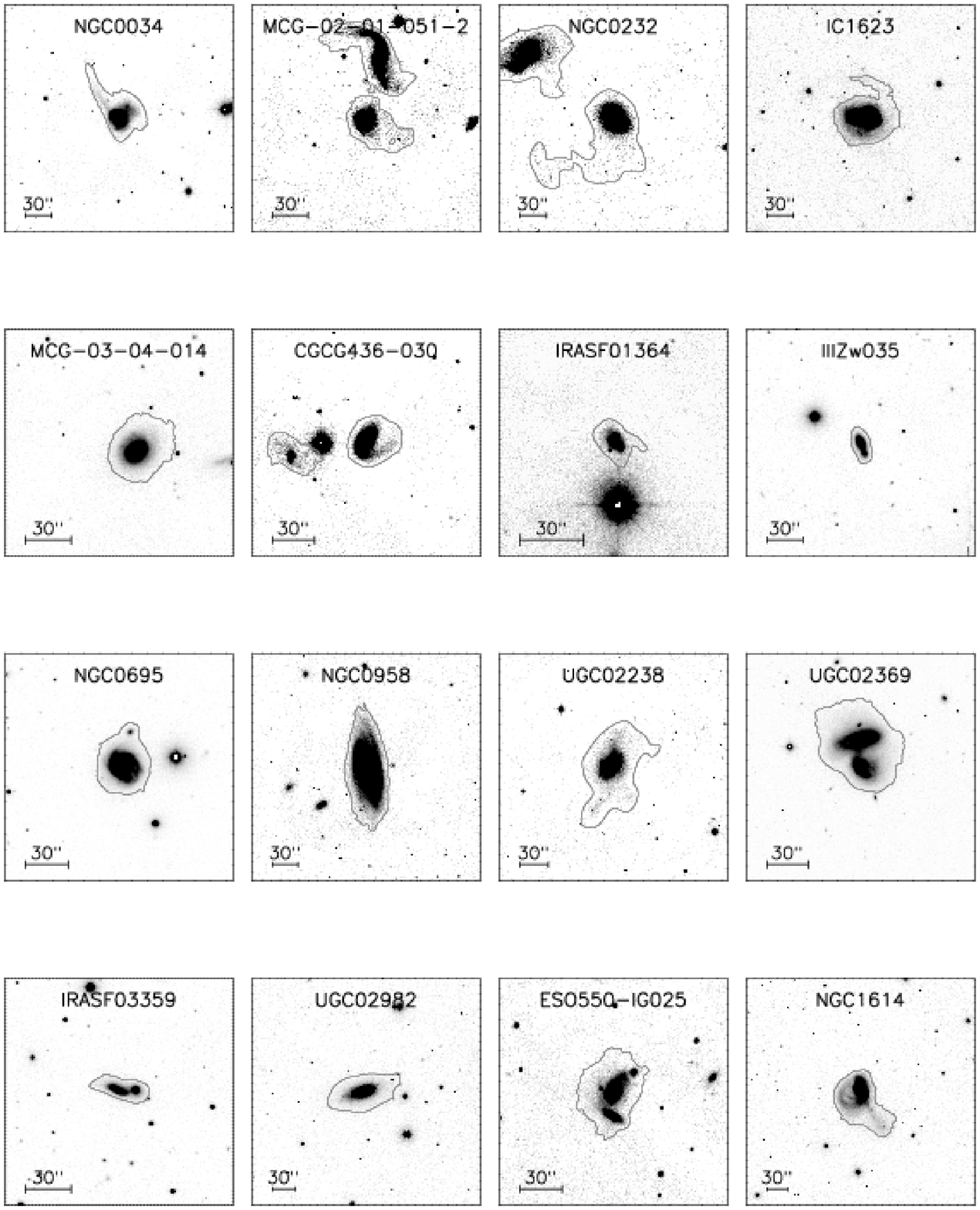}
 \end{figure}

\begin{figure}
 \centering
 \includegraphics[width=0.9\textwidth]{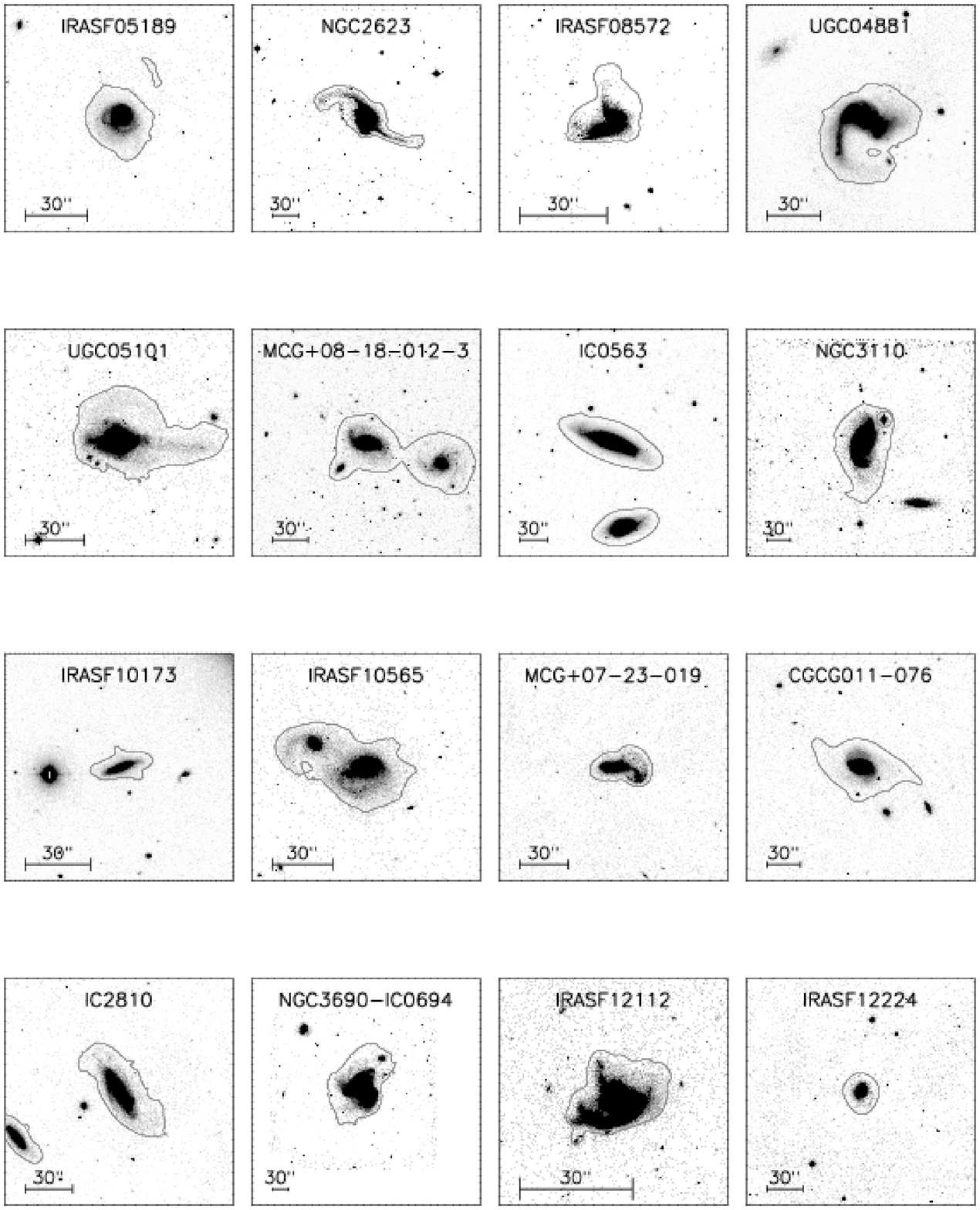}
 \end{figure}

\begin{figure}
 \centering
 \includegraphics[width=0.9\textwidth]{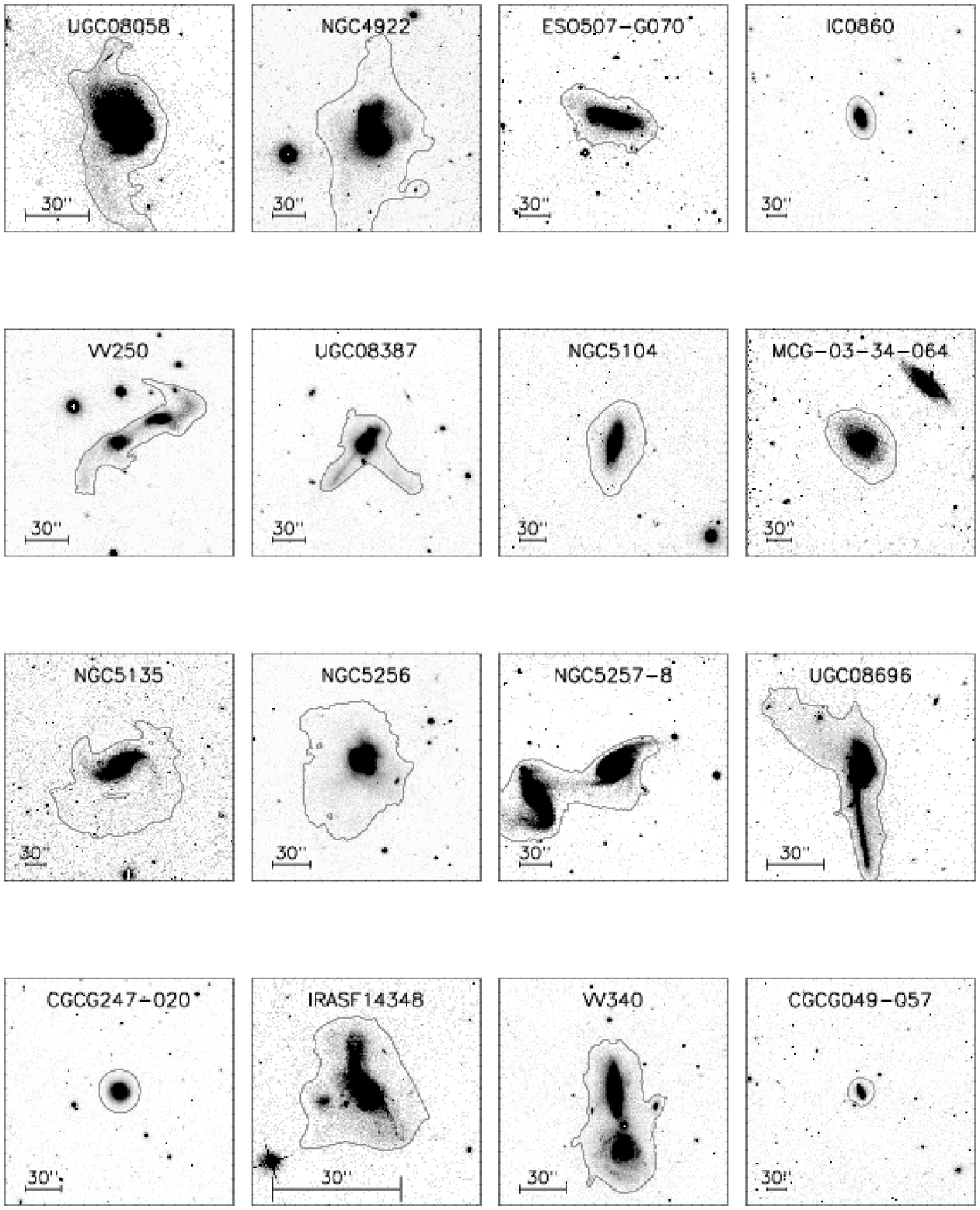}
 \end{figure}

  \begin{figure}
    \centering
    \includegraphics[width=0.99\textwidth]{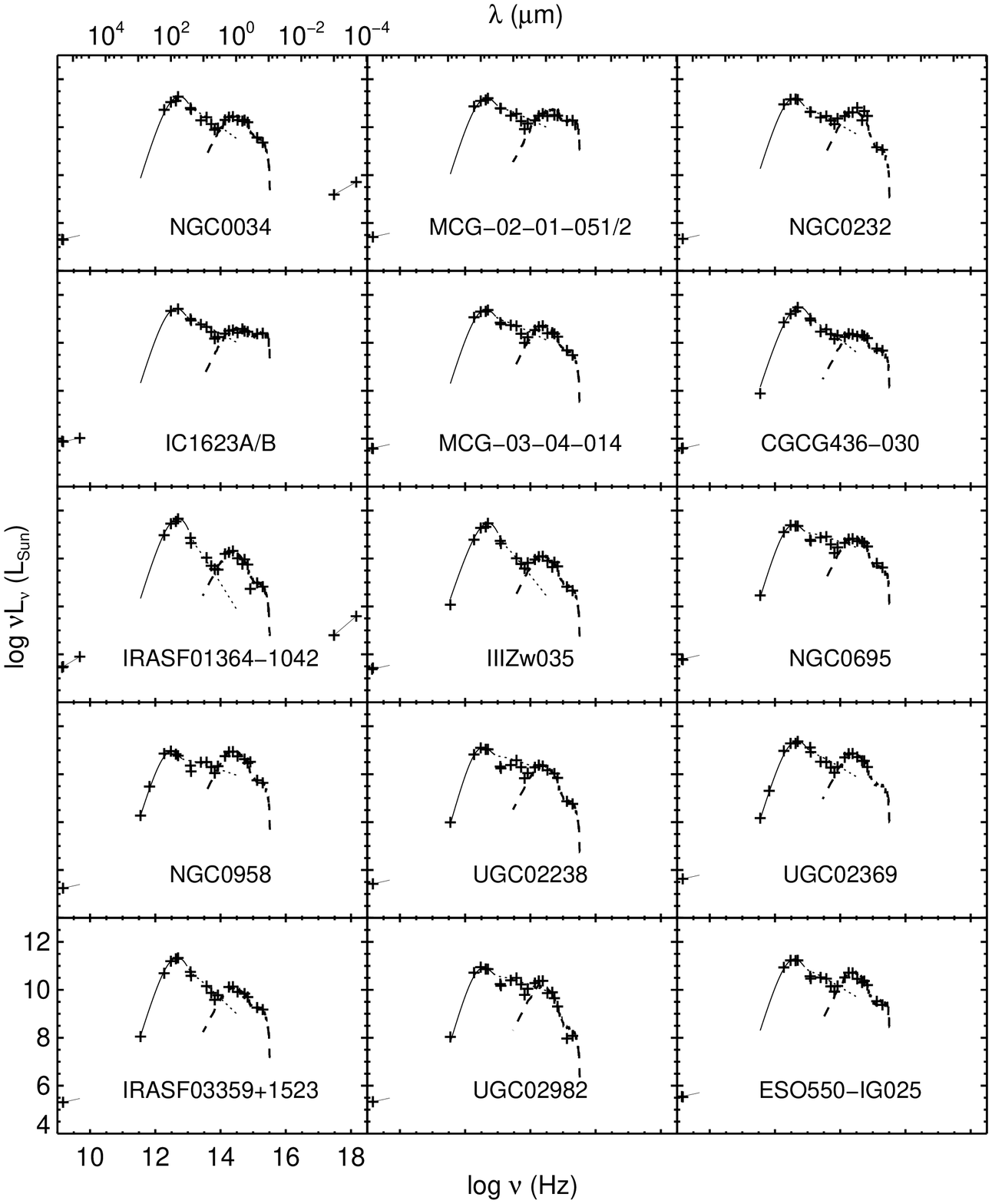}
  \end{figure}

  \begin{figure}
    \centering
    \includegraphics[width=0.99\textwidth]{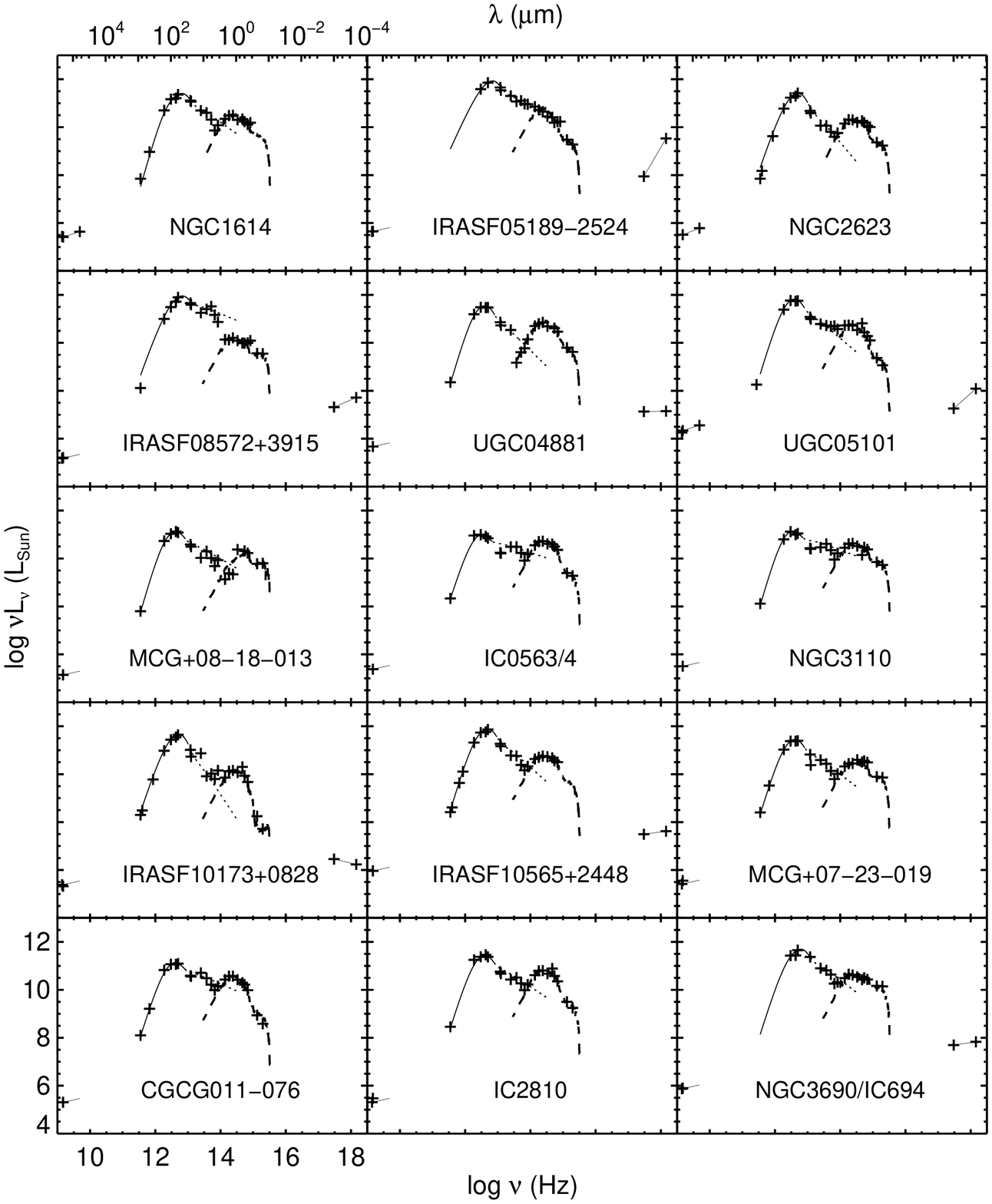}
  \end{figure}
  
  \begin{figure}
    \centering
    \includegraphics[width=0.99\textwidth]{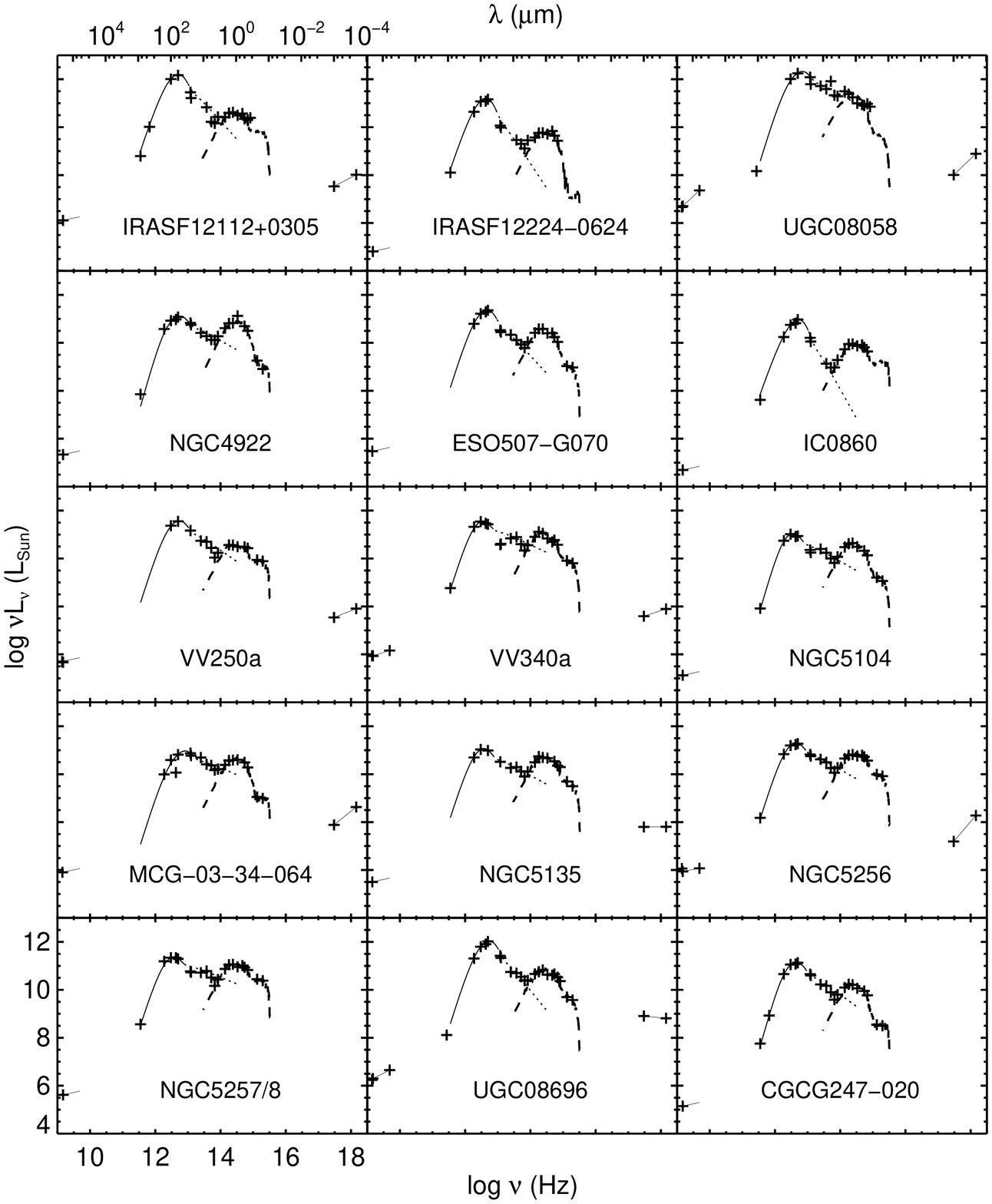}
  \end{figure}
  
  \begin{figure}
    \centering
    \includegraphics[width=0.99\textwidth]{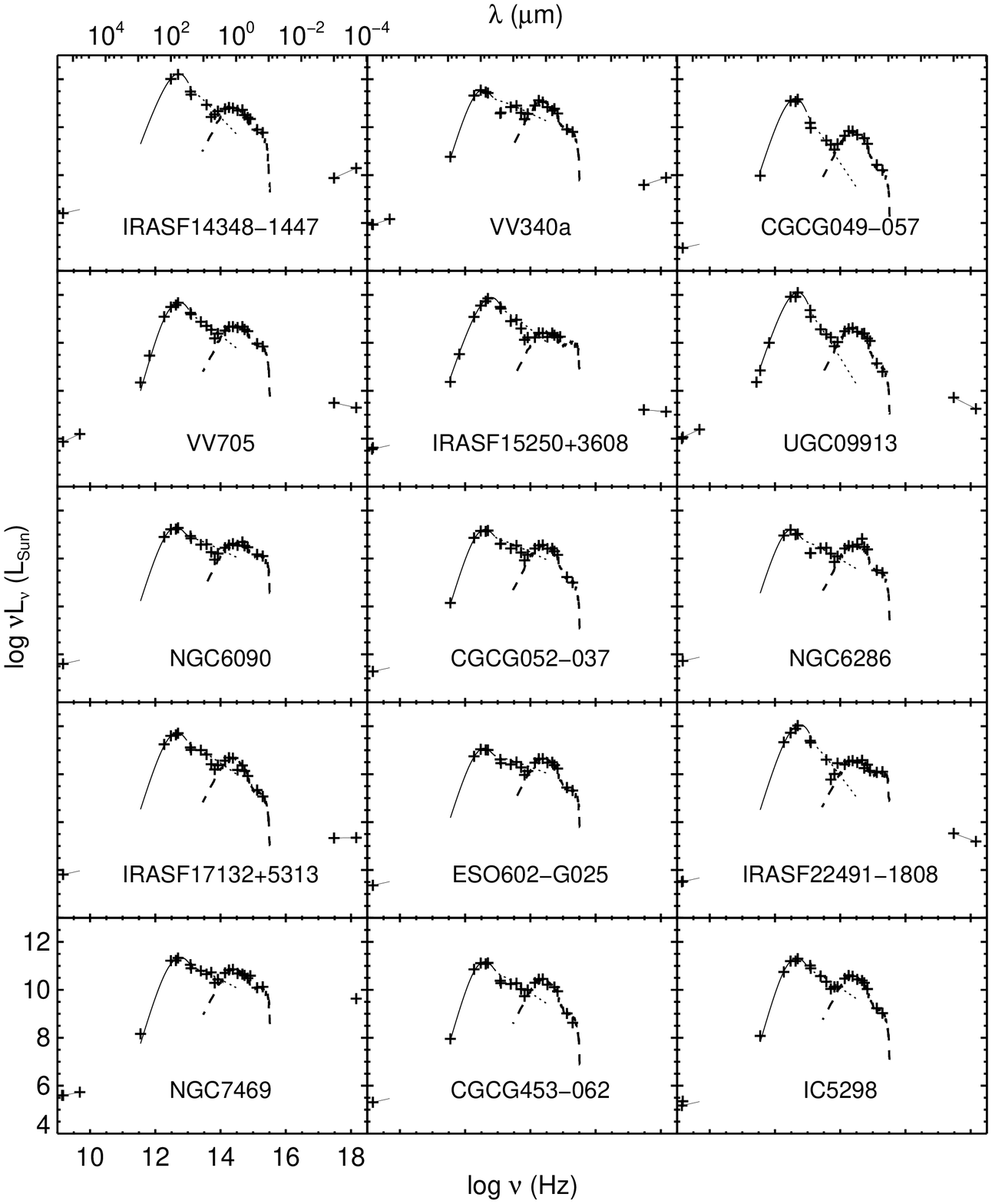}
  \end{figure}
  
\clearpage

\begin{deluxetable}{lcccrl}
    \centering
    \tabletypesize{\scriptsize}
    \tablewidth{0pt}
    \tablecolumns{6}
    \tablecaption{Local (U)LIRG Sample}
    \tablehead{    
      \colhead{RBGS Name} &  \colhead{RA} &  \colhead{Dec} & \colhead{log ($L_{\rm IR}$)$_{IRAS}$} & \colhead{$D_L$} &\colhead{Other Names}\\
        &\colhead{(J2000)} &\colhead{(J2000)} &\colhead{($L_{\odot}$)} & \colhead{(Mpc)} & \colhead{}
      	}
    \startdata
    NGC 0034          & 00 11 06.5  & $-$12 06 26  & 11.49 &  84.1 & NGC  0017, VV 860, Mrk 0938 \\
    MCG -02-01-051/2  & 00 18 50.4  & $-$10 22 08  & 11.48 & 117.5 & Arp 256, VV 352 \\
    NGC 0232          & 00 42 45.8  & $-$23 33 41  & 11.44 &  95.2 & VV 830 \\
    IC 1623A/B        & 01 07 47.2  & $-$17 30 25  & 11.71 &  85.5 & Arp 236, VV 114 \\
    MCG -03-04-014    & 01 10 08.9  & $-$16 51 10  & 11.65 & 144.0 &  \\
    CGCG 436-030      & 01 20 02.7  & $+$14 21 43  & 11.69 & 134.0 &  \\
    IRAS F01364-1042  & 01 38 52.9  & $-$10 27 11  & 11.85 & 210.0 &  \\
    III Zw 035        & 01 44 30.5  & $+$17 06 05  & 11.64 & 119.0 &  \\
\enddata
    \tablecomments{The $L_{\rm IR}$ column indicates infrared luminosities
      based on \emph{IRAS} measurements only.}
    \label{tbl:basicinfo}
  \end{deluxetable}

 \begin{deluxetable}{lcll}
    \centering
    \tabletypesize{\scriptsize}
    \tablewidth{0pt}
    \tablecolumns{4}
    \tablecaption{Major Data Sets}
    \tablehead{  
      \colhead{Telescope} & \colhead{No. Obj.}  & Bands & \colhead{Reference} \\
      }
    \startdata
$Chandra$ & 22  & 0.5-2keV, 2-10keV & \cite{Iwasawa11} \\ 
$GALEX$ & 57 & $NUV, FUV$ & \cite{Howell10} \\  
%HST & 56 & Evans et al. 2011 \\
UH2.2m & 64 & $B, V, I $& \cite{Ishida04} \\
UH2.2m & 8 & $U, B$ & \cite{Surace98,Surace00a,Surace00b} \\
UH2.2m & 46 & $R$ & Mazzerella et al., in prep. \\
2MASS & 64 & $J, H, K_{\rm s}$ & \cite{Skrutskie06,Jarrett03} \\
$Spitzer$ & 64 & 3.6,4.5,5.8,8.0,24,70,160 $\mu$m & Mazzarella et al., in prep. \\
$IRAS$ & 64 & 12, 25, 60, 100 $\mu$m & \cite{Sanders03} \\
JCMT/SCUBA & 42 & 450, 850 $\mu$m & \cite{Dunne00, Dunne01} \\
VLA & 64 & 1.49, 4.85 GHz & \cite{Condon90,Condon96} 
     \enddata
    \label{tbl:newobs}
  \end{deluxetable}

  \begin{deluxetable}{lcccl}
    \centering
    \tabletypesize{\scriptsize}
    \tablewidth{0pt}
    \tablecolumns{5}
    \tablecaption{New UH2.2m Observations}
    \tablehead{  
      \colhead{RBGS Name} & \colhead{Exp. Time}  & \colhead{Filter} & \colhead{Seeing} & \colhead{Obs. Date}\\
         & \colhead{(sec)} & \colhead{(Johnson)} & \colhead{(arcsec)} & \colhead{(UT)}
      }
    \startdata
    CGCG 436-030  & 180 & $R$ & $0\farcs7$ & 2008/09/01 \\
    IRAS F01364-1042  & 360 & $U$ & $0\farcs6$ & 2008/08/28 \\
    NGC 0695  & 180 & $R$ & $0\farcs7$ & 2008/09/01 \\
    IRAS F05189-2524  & 180 & $V$ & $0\farcs6$  & 2008/09/01 \\
                      & 180 & $I$ & $0\farcs6$ & 2008/09/01 \\
     IRAS F08572+3915  & 180 & $V$ & $0\farcs7$ & 2008/02/15 \\
     IRAS F10565+2448  & 330 & $U$ & $0\farcs6$ & 2008/02/18-19 \\
                        & 180 & $V$ & $0\farcs7$ & 2008/02/18 \\
     IRAS F12112+0305  & 180 & $V$ & $0\farcs5$ & 2008/02/19 \\
     UGC 08058  & 120 & $I$ & $0\farcs8$ & 2008/02/16 \\
     UGC 08696  & 120 & $V$ & $0\farcs5$ & 2008/02/17 \\
     IRAS F14348-1447  & 180 & $V$ & $0\farcs7$ & 2008/02/15 \\
     IRAS F22491-1808  & 180 & $V$ & $0\farcs7$ & 2008/08/28
     \enddata
    \label{tbl:uhobs}
  \end{deluxetable}

 \hspace{-0.5in}

\begin{landscape}
  \begin{deluxetable}{lcccccccccccc}
    \centering
    \tabletypesize{\scriptsize}
    \tablewidth{0pt}
    \setlength{\tabcolsep}{0.02in}
    \tablecolumns{13}
    \tablecaption{Local (U)LIRG Photometry (Jy; $\lambda~<~3\,\mu m$)}
    \tablehead{   % column headings
      \colhead{RBGS Name} &
      \colhead{$HX$} &
      \colhead{$SX$} &
      \colhead{$FUV$} &
      \colhead{$NUV$} &
      \colhead{$U$} &
      \colhead{$B$} &
      \colhead{$V$} &
      \colhead{$R$} &
      \colhead{$I$} &
      \colhead{$J$} &
      \colhead{$H$} &
      \colhead{$K_{\rm s}$} \\
      \colhead{} &
      \colhead{2-10 KeV} &
      \colhead{0.5-2 KeV} &
      \colhead{152.8 \AA} &
      \colhead{227.1 \AA} &
      \colhead{3655 \AA} &
      \colhead{4582 \AA} &
      \colhead{5377 \AA} &
      \colhead{6484 \AA} &
      \colhead{8570 \AA} &
      \colhead{1.241 $\mu$m} &
      \colhead{1.651 $\mu$m} &
      \colhead{2.165 $\mu$m} \\
      \colhead{} &
      \colhead{(2)} &
      \colhead{(3)} &
      \colhead{(4)} &
      \colhead{(5)} &
      \colhead{(6)} &
      \colhead{(7)} &
      \colhead{(8)} &
      \colhead{(9)} &
      \colhead{(10)} &
      \colhead{(11)} &
      \colhead{(12)} &
      \colhead{(13)}
      }
    \startdata
NGC 0034  & 1.586E-08$^a$ & 2.318E-08$^a$ & 5.249E-04 & 1.262E-03 & ... & 1.076E-02 & 1.629E-02 & 1.740E-02 & 2.679E-02 & 5.225E-02 & 6.529E-02 & 6.557E-02 \\
 & (...) & (...) & (6.480E-05) & (7.126E-05) & (...) & (5.949E-04) & (9.004E-04) & (1.122E-03) & (1.481E-03) & (1.357E-04) & (2.097E-04) & (1.783E-04) \\
MCG -02-01-051/2  & ... & ... & 2.255E-03 & 3.279E-03 & ... & 1.028E-02 & 1.368E-02 & ... & 2.032E-02 & 3.552E-02 & 3.808E-02 & 3.247E-02 \\
 & (...) & (...) & (2.784E-04) & (1.851E-04) & (...) & (1.136E-03) & (1.512E-03) & (...) & (2.246E-03) & (1.287E-04) & (3.044E-04) & (2.060E-04) \\
NGC 0232 & ... & ... & 2.038E-04 & 3.880E-04 & ... & 1.535E-02 & 3.071E-02 & 1.461E-02 & 6.925E-02 & 6.128E-02 & 8.037E-02 & 7.487E-02 \\
 & (...) & (...) & (1.538E-05) & (1.637E-05) & (...) & (2.097E-03) & (2.220E-03) & (9.417E-04) & (1.148E-02) & (9.594E-04) & (1.480E-03) & (1.448E-03) \\
IC 1623A/B  & ... & ... & 5.624E-03 & 7.898E-03 & ... & 1.871E-02 & 2.489E-02 & 3.569E-02 & 3.467E-02 & 6.292E-02 & 7.755E-02 & 7.716E-02 \\
 & (...) & (...) & (6.942E-04) & (4.459E-04) & (...) & (1.034E-03) & (1.375E-03) & (2.301E-03) & (1.916E-03) & (1.304E-04) & (1.906E-04) & (1.958E-04) \\
\enddata
\tablecomments{The photometry values have been measured
   with a uniform mask for each object.  \\ Exceptions in each column are marked by
   superscripted letters indicating literature or NED reference. \\
        Columns (2-3) X-ray Chandra-ACIS photometry \citep{Iwasawa11}\\
        Columns (4-5) NUV-FUV $GALEX$ photometry measured from image data in \cite{Howell10}\\
        Column (6) UV photometry from \cite{Surace00a} and this paper \\
        Column (7-8) B,V-band photometry from \cite{Ishida04,Surace98,Surace00b} \\
        Column (9) R-band photometry from Kim et al. (in preparation) \\
        Column (10) I-band photometry from \cite{Ishida04,Surace98,Surace00b} \\
        Columns (11-13) $J,H,K_{\rm s}$ photometry from  2MASS \\
        $^a$\,NED; $^b$\,\cite{Surace98,Surace00a,Surace00b}
        }
\label{tbl:photometry1}
\end{deluxetable}
\end{landscape}

\begin{landscape}
  \begin{deluxetable}{lccccccccccccccc}
    \centering
    \tabletypesize{\scriptsize}
    \tablewidth{0pt}
    \setlength{\tabcolsep}{0.01in}
    \tablecolumns{16}
    \tablecaption{Local (U)LIRG Photometry (Jy; $\lambda~>~3\,\mu m$)}
    \tablehead{   % column headings
      \colhead{Object} & 
      \colhead{IRAC1} & 
      \colhead{IRAC2} & 
      \colhead{IRAC3} & 
      \colhead{IRAC4} & 
      \colhead{IRAS1} &
      \colhead{MIPS1} & 
      \colhead{IRAS2} & 
      \colhead{IRAS3} & 
      \colhead{MIPS2} & 
      \colhead{IRAS4} & 
      \colhead{MIPS3} & 
      \colhead{SCUBA1} & 
      \colhead{SCUBA2} &
      \colhead{VLA1} & 
      \colhead{VLA2} \\ 
      \colhead{} & 
      \colhead{3.6$\mu$m} & 
      \colhead{4.5$\mu$m} & 
      \colhead{5.8$\mu$m} &
      \colhead{8.0$\mu$m} & 
      \colhead{12$\mu$m} & 
      \colhead{24$\mu$m} & 
      \colhead{25$\mu$m} & 
      \colhead{60$\mu$m} & 
      \colhead{70$\mu$m} & 
      \colhead{100$\mu$m} & 
      \colhead{160$\mu$m} & 
      \colhead{450$\mu$m} & 
      \colhead{850$\mu$m} & 
      \colhead{4.85GHz} & 
      \colhead{1.49GHz} \\
      \colhead{} &
      \colhead{(2)} &
      \colhead{(3)} &
      \colhead{(4)} &
      \colhead{(5)} &
      \colhead{(6)} &
      \colhead{(7)} &
      \colhead{(8)} &
      \colhead{(9)} &
      \colhead{(10)} &
      \colhead{(11)} &
      \colhead{(12)} &
      \colhead{(13)} &
      \colhead{(14)} &
      \colhead{(15)} &
      \colhead{(16)}
    }
    \startdata
NGC 0034  & 0.050 & 0.053 & 0.119 & 0.314 & 0.350 & 1.999 & 2.390 & 17.050 & 13.285 & 16.860 & 12.867 & ... & ... & ... & 0.061 \\ 
 & (0.005) & (0.005) & (0.005) & (0.005) & (0.032) & (0.100) & (0.055) & (0.045) & (1.993) & (0.135) & (3.860) & (...) & (...) & (...) & (0.001) \\ 
MCG -02-01-051/2  & 0.040 & 0.029 & 0.081 & 0.225 & 0.280 & 1.120 & 1.200 & 7.480 & 7.472 & 9.660 & 9.120 & ... & ... & ... & 0.041 \\ 
 & (0.004) & (0.003) & (0.004) & (0.004) & (0.034) & (0.056) & (0.055) & (0.048) & (1.121) & (0.138) & (2.736) & (...) & (...) & (...) & (0.001) \\ 
NGC 0232 & 0.098 & 0.071 & 0.129 & 0.276 & 0.360 & 1.223 & 1.280 & 10.050 & 12.341 & 17.140 & 16.965 & ... & ... & ... & 0.052 \\ 
 & (0.010) & (0.007) & (0.010) & (0.010) & (0.034) & (0.061) & (0.039) & (0.037) & (1.851) & (0.094) & (5.090) & (...) & (...) & (...) & (0.001) \\ 
IC 1623A/B  & 0.092 & 0.099 & 0.241 & 0.542 & 1.030 & 3.014 & 3.650 & 22.930 & ... & 31.550 & ... & ... & ... & 0.096 & 0.221 \\ 
 & (0.009) & (0.010) & (0.009) & (0.009) & (0.030) & (0.151) & (0.050) & (0.062) & (...) & (0.113) & (...) & (...) & (...) & (0.012) & (0.001) \\
\enddata
      \tablecomments{The photometry values have been measured
   with a uniform mask for each object.  Exceptions in each column are marked by
   superscripted letters indicating literature or NED reference. \\
        Columns (2-5) $Spitzer$-IRAC photometry from images in
        Mazzarella  et al., in prep.\\
        Columns (6) $IRAS$ photometry from \cite{Sanders03} \\
        Column (7) $Spitzer$-MIPS24 photometry from  images in
        Mazzarella et al., in prep.\\
        Column (8-9) $IRAS$ photmetry from \cite{Sanders03} \\
        Column (10) $Spitzer$-MIPS70 photometry from images in
        Mazzarella et al., in prep.\\
        Column (11) $IRAS$ photometry from \cite{Sanders03}  \\
        Column (12) $Spitzer$-MIPS160 photometry from images in
        Mazzarella et al., in prep.\\
        Columns (13-14) JCMT/SCUBA submillimeter photometry from \cite{Dunne00,Dunne01} \\
        Columns (15-16) VLA photometry from \cite{Condon90,Condon91}\\
          $^c$\,at 1.425 GHz -- photometry from \cite{Condon96}
        }     
   \label{tbl:photometry2}
   \end{deluxetable}
  \end{landscape}

  \begin{deluxetable}{ccccc}
    \centering
    \tabletypesize{\scriptsize}
    \tablewidth{0pt}
    \tablecolumns{3}
    \tablecaption{Mean SED Templates}
    \tablehead{   % column headings
      \colhead{log $\nu$} & 
      \colhead{$\lambda$} &
       \colhead{Band} & 
      \colhead{log $\nu{L_\nu}$} & 
      \colhead{log $\nu{L_\nu}$} \\
      \colhead{(Hz)} &
      \colhead{($\mu$m)} & 
       \colhead{(Name)} & 
      \colhead{ ($L_\odot$)} &
      \colhead{ ($L_\odot$)} \\
      \colhead{} &
     \colhead{} &
     \colhead{} &
      \colhead{(ULIRGs)} &
      \colhead{(LIRGs)}   
	}
    \startdata
18.16 & 2.07E-04 & HX     &  8.20 &  7.68 \\
17.48 & 9.93E-04 & SX     &  7.86 &  7.39 \\
15.29 &     0.15 & FUV    &  9.72 &  9.68 \\
15.11 &     0.23 & NUV    &  9.76 &  9.72 \\
14.91 &     0.37 & U      &  9.97 &  9.94 \\
14.82 &     0.46 & B      & 10.38 & 10.32 \\
14.75 &     0.54 & V      & 10.52 & 10.42 \\
14.67 &     0.65 & R      & 10.64 & 10.51 \\
14.54 &     0.86 & I      & 10.77 & 10.61 \\
14.38 &     1.24 & J      & 10.85 & 10.64 \\
14.26 &     1.65 & H      & 10.86 & 10.63 \\
14.14 &     2.17 & K      & 10.73 & 10.47 \\
13.93 &     3.56 & IRAC1  & 10.80 & 10.25 \\
13.82 &     4.51 & IRAC2  & 10.89 & 10.33 \\
13.72 &     5.76 & IRAC3  & 10.98 & 10.40 \\
13.58 &     7.96 & IRAC4  & 11.11 & 10.50 \\
13.41 &    11.60 & IRAS1  & 11.24 & 10.62 \\
13.10 &    23.84 & MIPS1  & 11.62 & 10.86 \\
13.10 &    23.88 & IRAS2  & 11.62 & 10.86 \\
12.69 &    61.49 & IRAS3  & 12.04 & 11.33 \\
12.62 &    72.56 & MIPS2  & 11.99 & 11.32 \\
12.47 &   101.98 & IRAS4  & 11.80 & 11.21 \\
12.28 &   156.96 & MIPS3  & 11.39 & 10.86 \\
11.82 &   449.57 & SCUBA1 &  9.89 &  9.33 \\
11.54 &   866.77 & SCUBA2 &  8.82 &  8.15 \\
 9.15 &  2.14E05 & VLA    &  6.19 &  5.57
   \enddata
    \label{tbl:templates}
  \end{deluxetable}

 \begin{deluxetable}{clllc}
    \centering
    \tabletypesize{\scriptsize}
    \tablewidth{0pt}
    \tablecolumns{4}
    \tablecaption{Spectral Index versus $L_{\rm IR}$}
    \tablehead{  
      \colhead{$\lambda$ range} & \colhead{$\alpha_j^i$}  & \colhead{$\bar{\alpha}$} & \colhead{$\bar{\alpha}$} & \colhead{$r$} \\
      \colhead{($\mu$m)} & \colhead{(index)} & \colhead{(LIRGs)} & \colhead{(ULIRGs)} &  
      }
    \startdata
$0.23-0.54$ & $\alpha_{0.23}^{0.54}$ & $-2.29\pm0.94$ & $-1.80\pm0.78$ &$+0.32$ \\ 
$2.2-4.5$ & $\alpha_{2.2}^{4.5}$ & $+1.49\pm0.70$ & $+0.84\pm0.69$ & $-0.30$ \\ 
$8.0-24.0$ & $\alpha_{8}^{24}$ & $-0.47\pm0.76$ & $-0.93\pm0.39$ & $-0.38$  
     \enddata
    \label{tbl:alphas}
  \end{deluxetable}

\begin{deluxetable}{cccccc}
    \centering
    \tabletypesize{\scriptsize}
    \tablewidth{0pt}
    \tablecolumns{6}
    \tablecaption{Variations in SEDs of (U)LIRGs}
    \tablehead{   % column headings
        \colhead{log $\nu$} & 
        \colhead{$\lambda$} &
        \colhead{Band} & 
        \colhead{$R$ } & 
        \colhead{$\sigma_R$} & 
        \colhead{$\Delta R$} \\
        \colhead{(Hz)} & 
        \colhead{($\mu$m)} & 
        \colhead{(Name)} & 
        \colhead{ } & 
        \colhead{ } &
        \colhead{ } 
    }
    \startdata
18.16 &   2.07E-04 & HX     &   -3.10   &  0.81   &  3.73  \\ 
17.48 &   9.93E-04 & SX     &   -3.69   &  0.54   &  2.24  \\ 
15.23 &       0.15 & FUV    &   -1.88   &  0.61   &  3.18  \\ 
15.11 &       0.23 & NUV    &   -1.83   &  0.56   &  2.88  \\ 
14.91 &       0.37 & U      &   -1.62   &  0.48   &  1.70  \\ 
14.82 &       0.46 & B      &   -1.23   &  0.47   &  1.68  \\ 
14.75 &       0.54 & V      &   -1.12   &  0.45   &  1.72  \\ 
14.67 &       0.65 & R      &   -1.03   &  0.44   &  1.76  \\
\enddata
  \tablecomments{SEDs normalized at 60$\mu$m for the complete sample of 53 LIRGs and 11 ULIRGS. Band names correspond to the filters used in Tables \ref{tbl:photometry1} and \ref{tbl:photometry2}.   $R$ corresponds to the ratio, log $[\frac{\nu L_\nu (\lambda)}{\nu L_\nu (60\mu{\rm m})}]$ .}
    \label{tbl:stat}
  \end{deluxetable}

  \begin{deluxetable}{lccc}
    \centering
    \tabletypesize{\scriptsize}
    \tablewidth{0pt}
    \tablecolumns{4}
    \tablecaption{Derived Infrared Luminosities, Dust Temperatures,
      and Dust Masses}
    \tablehead{
      \colhead{RBGS Name} &\colhead{log ($L_{\rm IR}$)$_{\rm mBB}$} &
      \colhead{$T_{\rm dust}$} & \colhead{log ($M_{\rm dust}$)} \\
         & \colhead{($L_{\odot}$)} & \colhead{($K$)} & \colhead{($M_{\odot}$)}
	}
    \startdata
             NGC0034     & 11.48$\pm$0.01   & 37.64$\pm$0.40   &  6.88$\pm$0.06 \\
     MCG-02-01-051/2     & 11.46$\pm$0.02   & 32.70$\pm$0.57   &  7.11$\pm$0.06 \\
             NGC0232     & 11.43$\pm$0.01   & 28.35$\pm$0.18   &  7.41$\pm$0.06 \\
           IC1623A/B     & 11.66$\pm$0.01   & 31.45$\pm$0.15   &  7.42$\pm$0.06 \\
       MCG-03-04-014     & 11.63$\pm$0.02   & 31.09$\pm$0.50   &  7.42$\pm$0.06 \\
         CGCG436-030     & 11.68$\pm$0.02   & 39.05$\pm$0.80   &  7.14$\pm$0.06 \\
     IRASF01364-1042     & 11.79$\pm$0.02   & 35.37$\pm$0.65   &  7.36$\pm$0.06 \\
            IIIZw035     & 11.62$\pm$0.02   & 35.86$\pm$0.65   &  7.35$\pm$0.06 \\
\enddata
    \label{tbl:dust}
  \end{deluxetable}

  \begin{deluxetable}{lcccc}
    \centering
    \tabletypesize{\scriptsize}
    \tablewidth{0pt}
    \tablecolumns{5}
    \tablecaption{Derived Stellar Masses}
    \tablehead{
      \colhead{RBGS Name} &\colhead{log ($M_{\rm fit}$)$_{\rm Sal}$} & \colhead{log ($M_{\rm H}$)$_{\rm Sal}$}  &
      \colhead{log ($M_{\rm fit}$)$_{\rm Cha}$} & \colhead{log ($M_{\rm H}$)$_{\rm Cha}$}  \\
         & \colhead{($M_{\odot}$)} & \colhead{($M_{\odot}$)}& \colhead{($M_{\odot}$)}& \colhead{($M_{\odot}$)} \\
      \colhead{} &
      \colhead{(2)} &
      \colhead{(3)} &
      \colhead{(4)} &
      \colhead{(5)} 
	}
    \startdata
   NGC0034  &  10.44  &  10.58  &  10.41  &  10.07  \\
   MCG-02-01-051/2  &   9.15  &  10.65  &  10.42  &  10.15  \\
   NGC0232  &  10.72  &  10.84  &  11.05  &  10.34  \\
   IC1623A/B  &   9.07  &  10.65  &  10.56  &  10.15  \\
   MCG-03-04-014  &  10.89  &  10.85  &  10.79  &  10.35  \\
   CGCG436-030  &   9.97  &  10.56  &   9.74  &  10.06  \\
   IRASF01364-1042  &  10.65  &  10.45  &  10.75  &   9.95  \\
   IIIZw035  &  10.15  &  10.25  &  10.26  &   9.75  \\
\enddata
     \tablecomments{Column (2) Mass fit (UV-opt SED) with Salpeter IMF  \cite[][]{Salpeter55} \\
        Column (3) Mass fit ($H$-band) with Salpeter IMF  \cite[][]{Salpeter55} \\
        Column (4) Mass fit (UV-opt SED) with Chabrier IMF  \cite[][]{Chabrier03} \\
        Column (5) Mass fit ($H$-band) with Chabrier IMF  \cite[][]{Chabrier03}
        }
    \label{tbl:masses}
  \end{deluxetable}

  \begin{deluxetable}{lrrrr}
    \centering
    \tabletypesize{\scriptsize}
    \tablewidth{0pt}
    \tablecolumns{5}
    \tablecaption{SFR and sSFR for Local (U)LIRGs}
    \tablehead{    
      \colhead{RBGS Name} &  
      \colhead{log SFR$_{\rm UV}$} & 
      \colhead{log SFR$_{\rm IR}$} & 
      \colhead{log SFR$_{\rm UV+IR}$} & 
      \colhead{log sSFR$_{\rm UV+IR}$} \\
      & \colhead{($M_{\odot}$ yr$^{-1}$)} &
      \colhead{($M_{\odot}$ yr$^{-1}$)} & 
      \colhead{($M_{\odot}$ yr$^{-1}$)} & 
      \colhead{(yr$^{-1}$)} 
      	}
    \startdata
   NGC0034  &   0.40  &   1.50  &   1.54  &  -8.88 \\ 
   MCG-02-01-051/2  &   0.38  &   1.50  &   1.53  &  -8.89 \\ 
   NGC0232  &   0.01  &   1.47  &   1.49  &  -9.51 \\ 
   IC1623A/B  &   0.49  &   1.72  &   1.74  &  -8.82 \\ 
   MCG-03-04-014  &   0.51  &   1.68  &   1.71  &  -9.10 \\ 
   CGCG436-030  &   0.19  &   1.71  &   1.72  &  -8.02 \\ 
   IRASF01364-1042  &  -0.16  &   1.80  &   1.81  &  -8.91 \\ 
   IIIZw035  &  -0.29  &   1.65  &   1.66  &  -8.61 \\
\enddata
    \label{tbl:sfr}
  \end{deluxetable}

\end{document}